\documentclass[prx,floatfix,superscriptaddress,amssymb,notitlepage,nofootinbib,longbibliography,twocolumn]{revtex4-2}
\usepackage{graphicx}
\usepackage{amsfonts}
\usepackage{amsmath}
\usepackage{amssymb}
\usepackage{comment}

\usepackage{lineno}
\usepackage{hyperref}
\hypersetup{
colorlinks=true,      % false: boxed links; true: colored links
linkcolor=cyan,         % color of internal links
citecolor=magenta,       % color of links to bibliography
filecolor=magenta,     % color of file links
urlcolor=cyan,          % color of external links
runcolor=cyan
}
\usepackage[dvipsnames]{xcolor}
\usepackage[capitalize]{cleveref}
\usepackage{soul}
\usepackage{bm}
\usepackage{threeparttable}
\usepackage{subfigure}
\usepackage{upgreek}
\usepackage{marginnote}
\usepackage{cancel}
\usepackage{afterpage}
\usepackage[normalem]{ulem}

\def \addCQuICPandAUNM {Center for Quantum Information and Control, Department of Physics and
  Astronomy, University of New Mexico, New Mexico, 87131, USA}

\def \addECEUNM {Department of Electrical and Computer Engineering, University of New Mexico,
  New Mexico, 87131, USA}

\newcommand{\beq}{\begin{equation}}
\newcommand{\eeq}{\end{equation}}
\newcommand{\beqnn}{\begin{equation*}}
\newcommand{\eeqnn}{\end{equation*}}
\newcommand{\bea}{\begin{eqnarray}}
\newcommand{\eea}{\end{eqnarray}}
\newcommand{\beann}{\begin{eqnarray*}}
\newcommand{\eeann}{\end{eqnarray*}}
\newcommand{\bes} {\begin{subequations}}
\newcommand{\ees} {\end{subequations}}

\newcommand{\braket}[2]{\langle #1 | #2\rangle}
\newcommand{\ket}[1]{ | #1\rangle}
\newcommand{\bra}[1]{\langle #1 | }

\newcommand{\ident}{\openone}
\newcommand{\Tr}{\mathrm{Tr}}

\newcommand{\ignore}[1]{}

\begin{document}
\title{Macroproperties vs. Microstates in the Classical Simulation of Critical Phenomena in Quench Dynamics of 1D Ising Models}

\author{Anupam Mitra}
\thanks{Corresponding author}
\email{anupam@unm.edu}
\affiliation{\addCQuICPandAUNM}

\author{Tameem Albash}
\affiliation{\addECEUNM}
\affiliation{\addCQuICPandAUNM}

\author{Philip Daniel Blocher}
\affiliation{\addCQuICPandAUNM}

\author{Jun Takahashi}
\affiliation{\addCQuICPandAUNM}

\author{Akimasa Miyake}
\affiliation{\addCQuICPandAUNM}

\author{Grant Biedermann}
\affiliation{Center for Quantum Research and Technology, Homer L. Dodge Department of Physics and Astronomy, University of Oklahoma, Oklahoma, 73019, USA}

\author{Ivan H. Deutsch}
\email{ideutsch@unm.edu}
\affiliation{\addCQuICPandAUNM}

\date{2024-12-14}

\begin{abstract}

We study the tractability of classically simulating critical phenomena in the quench dynamics of one-dimensional transverse field Ising models (TFIMs) using highly truncated matrix product states (MPS). We focus on two paradigmatic examples: a dynamical quantum phase transition (DQPT) that occurs in nonintegrable long-range TFIMs, and the infinite-time correlation length of the integrable nearest-neighbor TFIM when quenched to the critical point, where the quantities of interest involve equal time one- and two- point correlation functions, which we associate with macroproperties. For the DQPT, we show that the order parameters can be efficiently simulated with  heavy truncation of the MPS bond dimension. This can be used to reliably extract critical properties of the phase transition, including critical exponents, even when the full many-body state is not simulated with high fidelity. The long-time correlation length near the critical point is more sensitive to the full many-body state fidelity, and generally requires a large bond dimension MPS. Nonetheless, this can still be efficiently simulated with strongly truncated MPS because it can be extracted from the short-time behavior of the dynamics where entanglement is low. Our results provide illustrations of scenarios where accurate calculation of the full many-body state (microstate) is intractable due to the volume-law growth of entanglement, yet a precise specification of an exact microstate may not be required when simulating macroproperties that play a role in phases of matter of many-body systems. We also study the tractability of simulation using truncated MPS based on quantum chaos and equilibration in the models. We find a counterintuitive inverse relationship, whereby local expectation values are most easily approximated for chaotic systems whose exact many-body state is most intractable.
\end{abstract}

\maketitle

%
%%%%%%%%%%%%%%%%%%%%%%%%%%%%%%%%%%%%%%%%%%%%%%%%%%%%%%%%%%%%%%%%%%%%%%%%%%
\section{Introduction}
\label{sec:Introduction}
%%%%%%%%%%%%%%%%%%%%%%%%%%%%%%%%%%%%%%%%%%%%%%%%%%%%%%%%%%%%%%%%%%%%%%%%%%
%
Quantum simulation of many-body physics has long been considered to be a potential application where noisy intermediate scale quantum (NISQ) processors~\cite{preskill2018quantum} could exhibit a computational advantage over classical computers. While severely limited by control errors and decoherence, it is hoped that a NISQ device may still succeed when it probes ``universal properties'' of the model, characterized by local order parameters (typically one- or two-point correlation functions), that represent ``macroproperties'' of the system rather than ``microstates'' that depend on the exact many-body state. Macroproperties are more likely to be robust to small imperfections and to decoherence and hence yield a good approximation to the exact solution, in contrast to, e.g., Shor’s factoring algorithm \cite{Sho1997}, where even a single bit-flip can lead to a completely erroneous output. One thus may expect such quantum simulations to solve many-body physics problems without the overhead and complexity of a universal fault-tolerant quantum computer~\cite{Aha1997,Kni1998}.

A simulation task of particular interest is the nonequilibrium dynamics of closed quantum systems. Out-of-equilibrium systems are considered to be hard to simulate classically~\cite{cirac2012goals, eisert2015quantum, daley2022practical}, even for a one-dimensional chain of particles, because such systems can lead to a volume-law growth of  entanglement~\cite{nakagawa2018universality, qi2019measuring, alba2019quantum,aramthottil2023scrambling}. Central questions include the nature of quench dynamics and its relationship to universality~\cite{karrash2012luttinger, heyl2015scaling, nikoghosyan2016universality, mitra2018quantum}, thermalization~\cite{Kar2017, halimeh2017dynamical,halimeh2017prethermalization, zauner2017probing, Mi2022, panfil2023thermalization}, scrambling~\cite{Page1993, Hayden2007, Sekino2008, Liu2018}, quantum chaos~\cite{karthik2007entanglement, Hosur2016}, and transport properties~\cite{leviatan2017quantum, rakovszky2022dissipation, vonkeyserlingk2022operator, de2020superdiffusion, de2021stability, gopalakrishnan2022distinct, friedman2020diffusive, gopalakrishnan2019kinetic, de2022subdiffusive}. Numerous experiments have studied the quantum simulation of local order parameters in nonequilibrium dynamics based on analog emulations~\cite{byun2022finding, kim2022rydberg, zhang2017observationmany, de2023non, bernien2017probing, bluvstein2021controlling, bluvstein2022quantum, bohnet2016quantum, browaeys2020many, choi2023preparing, ebadi2021quantum,ebadi2022quantum, schauss2018quantum, scholl2021quantum, semeghini2021probing, weimer2010rydberg} and digital quantum circuits~\cite{noel2022measurement, Mi2022, niroula2023phase, hoke2023quantum, keenan2023evidence, kim2023evidence, kim2023evidence, kandala2019error}.

If quantum simulations of local order parameters are expected to be robust, it is natural to ask when such simulations are also computationally complex, in that they cannot be efficiently simulated on a classical computer. Indeed, as stated in Ref.~\cite{preskill2018quantum}, ``a major challenge for research using analog quantum simulators is identifying accessible properties of quantum systems which are robust with respect to error, yet are also hard to simulate classically.'' While high-fidelity classical simulation of the exact many-body state (the microstate) is generically intractable for sufficiently large systems and long times, this alone does not imply that a NISQ device will demonstrate a quantum advantage, since the complexity of simulating specific macroproperties may not be as difficult. Thus it is essential to devise classical algorithms that can approximate the desired output, rather than the entire many-body state, in order to understand where the boundary of quantum advantage occurs.

In this work, we revisit the complexity needed to approximate local expectation values in a particular class of one-dimensional (1D) nonequilibrium dynamics of Ising models. While such 1D models are far from where one expects the greatest quantum advantage, we focus on them here because they are a standard paradigm for quench dynamics, and their simplicity will help us to reveal the robustness of macroproperties versus microstates in quantum simulation. Our goal is to determine how the complexity of simulating the \emph{relevant output}, here specified by one- or two-point correlation functions, is different from the complexity of simulating the full many-body state. We focus, in particular, on quantitative estimates of universal properties, such as critical exponents, not previously studied.

To study the complexity of these tasks via classical simulation, we use matrix product state (MPS) tensor network simulation methods. Computational tractability is then measured by the entanglement that we need to retain in the simulation, as quantified by the MPS bond dimension. While high fidelity simulations of the full many-body state require a bond dimension that grows exponentially with system size at long times as the state becomes volume-law entangled, our approach is to determine the minimum bond dimension needed in the MPS simulation to accurately simulate the desired output. 

We study different cases to assess when quantities based on expectations of local observables are tractable with MPS simulations. Firstly, we consider a case where the quantities of interest are order parameters that involve space and time averaging, leading to some cancellation of errors. Secondly, we consider a situation where the quantities of interest are not space and time averaged order parameters, but specific two-point correlation functions, such as those that determine a correlation length. In this case, the accuracy of simulation may be expected to depend more strongly on the exact many-body state. We see a subtle relationship to integrability where chaos, scrambling, and many-body entanglement makes the estimation of expectation values of local observables easier due to effective high temperature thermalization. In the integrable case, while in principle estimating the long-time correlation length would require long-time evolution where volume-law entanglement makes MPS intractable, in practice the characteristic correlation length is already seen in short-time quenches. In all of these examples, a strongly truncated MPS simulation yields an excellent approximation, but for different reasons.

The remainder of the article is organized as follows. In Sec.~\ref{sec:ModelsAndMethods} we discuss a dynamical quantum phase transition (DQPT) occurring in 1D TFIM with variable-range interactions and review the different ways in which complexity arises in this model and its relation to chaos and the eigenstate thermalization hypothesis (ETH)~\cite{alba2015eigenstate, khodja2015relevance, d2016quantum, deutsch2018eigenstate,dymarsky2018subsystem, murthy2019bounds, noh2021eigenstate}. This sets the stage for understanding the conditions under which we expect simulating local order parameters to be numerically tractable. In Sec.~\ref{sec:DynamicalQuantumPhaseTransition} we study the estimation of critical properties near a $\mathbb{Z}_2$ symmetry breaking DQPT in the 1D TFIM with long-range interactions \cite{Zun2018,lang2018dynamical,halimeh2017dynamical,halimeh2017prethermalization,zauner2017probing, zhang2017observationmany, de2023non}. We provide numerical evidence to show that critical points and critical exponents can be well approximated using heavily truncated MPS. In Sec.~\ref{sec:CorrelationLengthFromQuench} we describe the estimation of long-time correlation lengths from short-duration quenches in the nearest-neighbor TFIM, as prescribed in~\cite{Kar2017}, and we relate the accuracy of estimates of the correlation length to the fidelity of the many-body state using truncated MPS. In Sec.~\ref{sec:RoleChaos} we explain the important role of quantum chaos and equilibration in the tractability of estimating local expectation values associated with quench dynamics. In Sec.~\ref{sec:SummaryOutlook} we summarize and provide an outlook for our results.

%
%%%%%%%%%%%%%%%%%%%%%%%%%%%%%%%%%%%%%%%%%%%%%%%%%%%%%%%%%%%%%%%%%%%%
\section{Models and methods}
\label{sec:ModelsAndMethods}
%%%%%%%%%%%%%%%%%%%%%%%%%%%%%%%%%%%%%%%%%%%%%%%%%%%%%%%%%%%%%%%%%%%%
%
%
%%%%%%%%%%%%%%%%%%%%%%%%%%%%%%%%%%%%%%%%%%%%%%%%%%%%%%%%%%%%%%%%%%%%
\subsection{Models}
\label{sec:Models}
%%%%%%%%%%%%%%%%%%%%%%%%%%%%%%%%%%%%%%%%%%%%%%%%%%%%%%%%%%%%%%%%%%%%
%
In the next two sections on critical phenomena, we focus on 1D TFIMs with variable-range interactions along the $z$-axis characterized by a power-law with exponent $\alpha \geq 0$ and a transverse field along the $x$-axis with magnitude $B$. For $n$ qubits the Hamiltonian is
\begin{equation} 
\begin{aligned}
  H = &
  - \frac{J_0}{\mathcal{N}} \sum_{\ell_1 < \ell_2}
  \frac{\sigma^z_{\ell_1} \sigma^z_{\ell_2}}
       {|\ell_1 - \ell_2|^{\alpha}}
  - B \sum_{\ell} \sigma^x_{\ell}
  ,
  \label{eq:HamiltonianPowerLawTFIM}
\end{aligned}
\end{equation}
where $\mathcal{N}$ is the Kac normalization factor~\cite{halimeh2017prethermalization,
halimeh2017dynamical, zhang2017observationmany, de2023non, Zun2018, lang2018dynamical}
\begin{equation}
  \mathcal{N}(\alpha, n) = \frac{1}{n-1}\sum_{\ell_1 < \ell_2} \vert \ell_1 -
  \ell_2 \vert^{-\alpha}
  = \frac{1}{n-1}\sum_{\ell=1}^{n-1} (n-\ell)\ell^{-\alpha}.
  \label{eq:KacNormalization}
\end{equation}
Such models have been implemented in atomic platforms, including trapped ions with $0 \leq \alpha \leq 3$~\cite{foss2013nonequilibrium, foss2013dynamical, bohnet2016quantum, jurcevic2017direct, zhang2017observationmany, de2023non, zhang2017observationdiscrete, monroe2021programmable}. Extensions of this model with a longitudinal field have been implemented in Rydberg atom arrays with $\alpha = 6$~\cite{bernien2017probing,weimer2010rydberg, schauss2018quantum,browaeys2020many, scholl2021quantum,ebadi2021quantum, bluvstein2021controlling}. Experiments have studied quench and driven dynamics \cite{bohnet2016quantum, zhang2017observationmany, de2023non,bernien2017probing, schauss2018quantum,bluvstein2021controlling, bluvstein2022quantum, choi2023preparing} as well as quasi-adiabatic passages~\cite{bernien2017probing, ebadi2021quantum,scholl2021quantum, semeghini2021probing,ebadi2022quantum, kim2022rydberg, byun2022finding} to probe ground states in exotic quantum phases of matter~\cite{bernien2017probing, ebadi2021quantum, scholl2021quantum, semeghini2021probing, ebadi2022quantum} and approximate solutions to graph problems encoded in Ising models~\cite{ebadi2022quantum, kim2022rydberg, byun2022finding}.

For $\alpha = 0$, this is a fully connected model, and the Hamiltonian governs the integrable dynamics of the collective spin vector $\mathbf{S} = \sum_{\ell} \boldsymbol{\sigma}_{\ell}/2$ whose total angular momentum $\mathbf{S}^2$ is conserved. This case of $\alpha = 0$ is an example of a Lipkin-Meshkov-Glick (LMG) model~\cite{Lip1965}. Here the thermodynamic limit is the mean-field limit, with all dynamics described by the motion of a unit vector on a spherical phase space~\cite{munoz2020simulation,chinni2021effect, munoz2021nonlinear, chinni2022trotter, munoz2022floquet, munoz2023phase, munoz2020simulation}. When $\alpha \to \infty$ one obtains the integrable 1D nearest-neighbor TFIM~\cite{lieb1961two, pfeuty1970one, pfeuty1971ising, derzhko1998numerical, sengupta2004quench, calabrese2011quantum, calabrese2012quantum1, calabrese2012quantum2}. Simulating dynamics of this model is classically tractable using the Onsager algebra of strings of local spin operators~\cite{lychkovskiy2021closed} or using a transformation to Gaussian Fermionic Hamiltonians~\cite{lieb1961two, pfeuty1970one, pfeuty1971ising, derzhko1998numerical, sengupta2004quench, calabrese2011quantum, calabrese2012quantum1, calabrese2012quantum2}. For all other values of $\alpha$ the system is generally nonintegrable, except for fine-tuned values of $B$ and $J_0$. We explore the integrability of these models through an analysis of level spacing statistics in Appendix~\ref{app:EnergyLevelSpacing}. Our goal is to study how well we can approximate one- and two-point correlations by truncating the bond dimension to some maximum value $\chi_{\max}$ where classical simulation is tractable.

We consider quench dynamics where all of the spins are initially polarized along some axis (a spin coherent state). The initial product state evolves according to the Hamiltonian in Eq.~\eqref{eq:HamiltonianPowerLawTFIM}, generally leading to volume-law entanglement, including for most integrable models~\cite{calabrese2005evolution, schollwock2011density, zhang2017observationmany, de2023non, bernien2017probing}, except in some special cases~\cite{lerose2020origin,lerose2020bridging}. Small values of $\alpha \lesssim 0.5$ typically yield Hamiltonians close to the collective spin model that have slow growth of entanglement~\cite{lerose2020origin, lerose2020bridging}, while large values of $\alpha \gtrsim 6$ lead to Hamiltonians close to the one-dimensional nearest-neighbor model that have fast growth of entanglement~\cite{calabrese2005evolution}. The former case, with long-range interactions, is amenable to analyses using perturbations to the collective spin model, while the latter case with short-range interactions is amenable to 1D tensor networks. Nevertheless, it is not apriori obvious whether either approach would be apropos for intermediate values of $1 \leq \alpha \leq 2$. Since we are considering 1D systems, we use matrix product states (MPS) with bounded entanglement content to represent the quantum state~\cite{perez2006matrix, schollwock2011density, vidal2003efficient,vidal2004efficient, daley2004time, white2004real, haegeman2011time, haegeman2016unifying,vanderstraeten2019tangent, zhou2020limits}. We review MPS representations in the next subsection.

The degree of chaos and ergodicity generated by dynamics determines the degree to which there is many-body ``scrambling"~\cite{Furuya1998, Lakshminarayan2001, Hayden2007, Sekino2008, Liu2018}. Due to the growth of entanglement associated with scrambling, the ETH implies that the expectation values of local observables at sufficiently long times are well-approximated using thermal expectation values, with the temperature determined by the initial energy of the system~\cite{deutsch2018eigenstate, d2016quantum, mori2018thermalization}. The exact time needed for equilibration depends on the Hamiltonian and its associated degree of chaos, the initial state, and other details of the model~\cite{d2016quantum, mori2018thermalization}. The most extreme equilibration occurs when the dynamics are maximally scrambling, and the many-body state has statistical properties of a typical state in Hilbert space~\cite{Leb1993, Gol2006, Leb2007, Bar2009, Dub2012, San2012, Fac2015, d2016quantum, Mi2022} with near maximum entanglement~\cite{Page1993, Foong1994, Sanchez1995, Sen1996}. In that case the local reduced states are the infinite temperature states, or maximally mixed states of appropriate dimension. We expect that the most interesting circumstances for quantum advantage in quantum simulation would correspond to those cases where a sufficient amount of entanglement is generated to make classical simulation of the full many-body state intractable, but not enough to render the local reduced states close to the maximally mixed state.

In the following two sections, we will focus on the problem of simulating critical properties of the TFIM. Models described by Eq.~\eqref{eq:HamiltonianPowerLawTFIM} have a ground-state quantum phase transition around $B/J_0 \approx 1$~\cite{Kar2017, halimeh2017dynamical, halimeh2017prethermalization, Zun2018,Tit2019}. For long-range interactions ($0\lesssim \alpha \lesssim 2$), the ground-state quantum phase transition survives as a thermal equilibrium phase transition at nonzero temperatures. Its properties can be accessed through long-time-averaged expectation values of local observables after a quench. The time and space averaged expectation values show a phase transition, which is considered a ``dynamical quantum phase transition" (DQPT)~\cite{halimeh2017dynamical, halimeh2017prethermalization, zhang2017observationmany, de2023non, Zun2018}, which we will study in Sec~\ref{sec:DynamicalQuantumPhaseTransition}. For short-range interactions ($\alpha \gtrsim 2$) this is not the case; the thermal phase transition and DQPT are absent. Nevertheless, signatures of the ground-state quantum phase transition persist, such as the critical behavior of the short-range spin-spin correlations near the critical point~\cite{Kar2017, Tit2019}, as we will study in Sec.~\ref{sec:CorrelationLengthFromQuench}. These DQPTs are related to DQPTs occurring due to nonanalytic behavior of a Loschmidt echo amplitude ~\cite{jurcevic2017direct, halimeh2017dynamical, halimeh2017prethermalization, Zun2018}. Moreover the DQPTs arising due to nonanalyticies in the Loschmidt amplitudes are due to both avoided crossing in the entanglement spectrum and qualitative changes in the overlap of the time evolved state with the initial state~\cite{de2021entanglement}. In all cases, thermal and similar states can be effectively accessed through quench dynamics.

MPS have been used extensively to study equilibrium and non-equilibrium properties of TFIMs~\cite{nagaj2008quantum, banuls2009matrix, koffel2012entanglement, wall2015out, pang2019critical, hashizume2022dynamical, kaicher2023mean, suzuki2012quantum, jaschke2017critical, jaschke2018open1, jaschke2018open2, Zun2018, halimeh2017dynamical, halimeh2017prethermalization, de2021entanglement, ran2020encoding, fossfeig2022entanglement}. While the 1D nearest neighbor TFIM can be solved by mapping to free fermions, it has been studied using MPS to describe and benchmark MPS algorithms~\cite{jaschke2018open1, jaschke2018open2}. Properties of the ground state of long-range interacting models have been studied using MPS~\cite{koffel2012entanglement}, as well as of non-equilibrium critical phenomena in TFIMs~\cite{jaschke2017critical, Zun2018, de2021entanglement}. In Ref.~\cite{jaschke2017critical}, the authors used MPS to study the density of defects in a linear sweep of the TFIM Hamiltonian. In Ref.~\cite{Zun2018}, the authors used MPS to study two different kinds of dynamical quantum phase transitions in TFIMs -- one based on the nonanalytic behavior of the Loschmidt echo, and another based on a long-time and space averaged order parameter. The latter of these will be the focus of Sec.~\ref{sec:DynamicalQuantumPhaseTransition}. The former has been studied with MPS, to study prethermalization ~\cite{halimeh2017prethermalization}, prepare a phase diagram of the DQPT~\cite{halimeh2017dynamical} and establish a relationship between the entanglement spectrum and the nonanalyticity in the Loschmidt echo~\cite{de2021entanglement}. MPS ansatz for ground states of the TFIM have also been encoded into quantum circuits~\cite{ran2020encoding}. Moreover, MPS approximations of the ground state of the TFIM have been prepared on a trapped ion quantum computer to study critical phenomenon as an early demonstration of quantum simulation using quantum computers~\cite{fossfeig2022entanglement}. We review MPS in the next subsection.

%
%%%%%%%%%%%%%%%%%%%%%%%%%%%%%%%%%%%%%%%%%%%%%%%%%%%%%%%%%%%%%%%%%%%%%%%%%%%%%%%%
\subsection{Matrix Product States: Representations and Dynamics}
\label{app:MatrixProductStates}
%%%%%%%%%%%%%%%%%%%%%%%%%%%%%%%%%%%%%%%%%%%%%%%%%%%%%%%%%%%%%%%%%%%%%%%%%%%%%%%%
%

Since we are considering one-dimensional systems, we use MPS to represent the quantum state~\cite{perez2006matrix, schollwock2011density, vidal2003efficient,vidal2004efficient, daley2004time, white2004real, haegeman2011time, haegeman2016unifying,vanderstraeten2019tangent, zhou2020limits}. We employ open boundary conditions, hence the MPS representation of the $n$ qubit system is written as
\begin{equation}
\begin{aligned}
  \ket{\psi} 
  = \sum_{z \in \left\{\uparrow, \downarrow \right\}^n} 
  \sum_{j_1, \dots, j_{n-1}} 
  &
  (\mathcal{M}_1^{z_1})_{ j_1}
  (\mathcal{M}_2^{z_2})_{j_1 j_2}\dots(\mathcal{M}_n^{z_n})_{j_{n-1}} 
  \\ & 
  \ket{z_1} \otimes \ket{z_2} \otimes \ket{z_n} \ ,
  \label{eq:MatrixProductState}
\end{aligned}
\end{equation}
where $\left\{ \mathcal{M}_{\ell}^{z_\ell}\right\}$ are up to $\chi \times \chi$ matrices (the subscript $\ell$ labels the position, and the superscript $z_\ell$ labels the basis state). For open boundary conditions $\mathcal{M}_{1}^{z_1}$ and $\mathcal{M}_{n}^{z_n}$ are row and column vectors, respectively. The ``bond dimension" $\chi \le 2^{\lfloor n/2 \rfloor}$, and $\ket{z}$ is an $n$-qubit computational basis state~\cite{perez2006matrix, schollwock2011density, vidal2003efficient,vidal2004efficient, daley2004time, white2004real, haegeman2011time, haegeman2016unifying,vanderstraeten2019tangent}. While this representation can be used for an arbitrary quantum state with $\chi$ potentially scaling exponentially with $n$, it is useful for classical simulation when it can efficiently approximate the desired quantum state, meaning that the bond dimension $\chi$ scales polynomially with $n$~\cite{vidal2003efficient, vidal2004efficient}. The maximum entanglement content of a quantum state expressed in MPS form with bond dimension $\chi$ is related to $\chi$ via
\begin{equation}
    \mathcal{S}_0 = \ln(\chi),
    \label{eq:Renyi0EntropyBondDimension}
\end{equation}
where $\mathcal{S}_0$ is the R{\'e}nyi-0 (or Hartley) entanglement entropy of half of the spin array with the other half. All other R{\'e}nyi-$k$ entropies are bounded by the the R{\'e}nyi-0 entropy as $\mathcal{S}_k \leq \mathcal{S}_0$. Throughout this article we measure entropies in nats. The bond dimension thus acts as an entanglement measure for our classical simulation budget. Such a truncation may lead to a poor approximation to the full many-body state (microstate) but an excellent approximation to macroproperties.

To simulate dynamics for nearest-neighbor models in Sec.~\ref{sec:CorrelationLengthFromQuench} and Sec.~\ref{sec:RoleChaos} we employ a 4th order time-evolving block decimation (TEBD) method~\cite{vidal2003efficient,vidal2004efficient, daley2004time, white2004real}, and for long range interacting models in Sec.~\ref{sec:DynamicalQuantumPhaseTransition}, we use the time-dependent variational principle (TDVP) method~\cite{haegeman2011time, haegeman2016unifying, vanderstraeten2019tangent,TimeEvoMPS}. In both cases, we start with an initial spin coherent state in the MPS form, Eq.~\eqref{eq:MatrixProductState}. During each time step we truncate the resultant MPS using a singular value decomposition (SVD) to bond-dimension $\chi_{\max}$. In the main text, for brevity, we drop the subscript with the implicit assumption $\chi \equiv \chi_{\max}$. Both TEBD and TDVP have runtimes that scale polynomially in the system size $n$ and bond dimension $\chi$ as $\mathcal{O}(n \chi^3)$~\cite{vidal2004efficient, daley2004time, haegeman2011time, vanderstraeten2019tangent}. Therefore, being able to calculate quantities of interest to the desired precision using TEBD or TDVP with a polynomially scaling (with system size and time) $\chi$ implies an efficient classical simulation, regardless of the classical tractability of the full many-body wave function.

%
%%%%%%%%%%%%%%%%%%%%%%%%%%%%%%%%%%%%%%%%%%%%%%%%%%%%%%%%%%%%%%%%%%%%%%%%%%%%%%%%
\subsection{Local observables and local reduced states}
\label{app:LocalReducedStates}
%%%%%%%%%%%%%%%%%%%%%%%%%%%%%%%%%%%%%%%%%%%%%%%%%%%%%%%%%%%%%%%%%%%%%%%%%%%%%%%%
%
The expectation value of a local observable $\mathcal{A}$ is determined by the reduced density operator $\rho$ of the spins participating non-trivially in the observable~$\mathcal{A}$. Two simulations using different maximum bond dimension $\chi_1$ and $\chi_2$ typically give different density operators $\rho_1$ and $\rho_2$, and we can consider the squared difference in expectation value of the observable~$\mathcal{A}$ for the two simulations. This is upper-bounded by
\begin{eqnarray}
    \Tr\left(\left(\rho_1 - \rho_2\right) \mathcal{A}\right)^2 
    & \leq &  
    \Tr\left(\mathcal{A}^\dagger \mathcal{A}\right) 
    \Tr\left(\left(\rho_1 - \rho_2\right)^2\right) 
    \nonumber \\
    & = & 
    \Tr\left(\mathcal{A}^\dagger \mathcal{A}\right) \,
    \mathcal{D}_{\mathrm{HS}}^2 \left(\rho_1, \rho_2\right),
    \label{eq:ExpectationValueBoundHSDistance}
\end{eqnarray}
where
\begin{equation}
    \mathcal{D}^2_{\mathrm{HS}}(\rho_1, \rho_2)  \equiv \Tr \left((\rho_1 - \rho_2)^\dagger (\rho_1 - \rho_2) \right)
    \label{eq:HSDistanceDefinition}
\end{equation}
is the squared Hilbert-Schmidt (HS) distance between the operators $\rho_1$ and $\rho_2$ (see Appendix~\ref{app:Bounds} for additional details). Since multiple many-body pure states are consistent with the same reduced density operator, we consider the former to be microstates and the latter to correspond to macroproperties.

%
%%%%%%%%%%%%%%%%%%%%%%%%%%%%%%%%%%%%%%%%%%%%%%%%%%%%%%%%%%%%%%%%%%%%%%%%%%%%%%%%
\section{Simulating criticality in a dynamical quantum phase transition}
\label{sec:DynamicalQuantumPhaseTransition}
%%%%%%%%%%%%%%%%%%%%%%%%%%%%%%%%%%%%%%%%%%%%%%%%%%%%%%%%%%%%%%%%%%%%%%%%%%%%%%%%
%
As a first example we simulate a DQPT associated with long-range interacting one-dimensional TFIMs as in Eq.~\eqref{eq:HamiltonianPowerLawTFIM}. The phases can be characterized by a dynamical order parameter given by the long-time average of expectation values of local observables. The order parameter characterizes two phases where the $\mathbb{Z}_2$ symmetry remains broken and is dynamically restored, respectively~\cite{Yuz2006, Sci2010,zhang2017observationmany, de2023non, Zun2018, lang2018dynamical}. Phenomenology associated with this order parameter description has been studied theoretically in Refs.~\cite{halimeh2017dynamical, halimeh2017prethermalization, zauner2017probing, lang2018dynamical, Zun2018} and observed in a 53-qubit ion-trap quantum simulator~\cite{zhang2017observationmany, de2023non}. 

To observe the DQPT we initialize the state as $\vert \psi (0) \rangle = \vert\hspace{-0.1cm} \uparrow_z \rangle^{\otimes n}$, corresponding to one of the two degenerate ground states of the Hamiltonian with $B/J_0 = 0$~\cite{zhang2017observationmany, de2023non, Zun2018, lang2018dynamical}. This choice explicitly breaks the $\mathbb{Z}_2$ symmetry of the system associated with a global $\pi$-rotation about the $x$-axis. We then quench to different finite values of $B/J_0$ and let the system evolve accordingly. For $\alpha \lesssim 2$ there exists a critical value of the quench parameter $B/J_0$ that separates two different phases of the asymptotic long-time quantum state in the thermodynamic limit~\cite{Zun2018,lang2018dynamical,halimeh2017dynamical}, corresponding to the $\mathbb{Z}_2$ symmetry remaining broken ($B/J_0 < (B/J_0)_{\mathrm{c}}$) and being restored ($B/J_0 > (B/J_0)_{\mathrm{c}}$). A dynamical order parameter used to probe this phase transition is the time-averaged $z$-magnetization
\begin{eqnarray}
    M_{z}(n, t)&=& \frac{1}{t} \frac{1}{n} \sum_{\ell=1}^n \int_{0}^{t} \mathrm{d} t'  \langle \psi(t') \vert \sigma_{\ell}^{z} \vert \psi(t') \rangle \nonumber  \\ 
    &\equiv&    \frac{1}{t} \frac{1}{n}  \int_{0}^{t} \mathrm{d} t' \langle 2S_{z} \rangle (t') ,
  \label{eq:MzOrderParameter}
\end{eqnarray}
which, in the thermodynamic limit $n \to \infty$ and long-time limit $J_0 t \to \infty$, is zero in the symmetry restored phase and nonzero in the symmetry broken phase.

A semiclassical picture for this behavior is the following. The two ferromagnetic states can be thought of as degenerate local minima of the energy landscape, and they are separated by an energy barrier. For $B/J_0 < (B/J_0)_{\mathrm{c}}$ the time-evolved state remains confined to one side of the barrier and tunneling to the other side is suppressed in the thermodynamic limit, giving rise to a time-averaged non-zero magnetization. For $B/J_0 > (B/J_0)_{\mathrm{c}}$ the state is above the separatrix orbit and is no longer confined by the energy barrier to a single side. Oscillations occur between the two ferromagnetic states, resulting in a vanishing time-averaged magnetization in the long-time limit~\cite{chinni2021effect, munoz2020simulation}. We elaborate further on this semiclassical picture using the case of $\alpha = 0$ in Appendix~\ref{app:DQPTSimplePicture}.

A more convenient probe of the phase transition is the long-time averaged two-spin correlation,
\begin{eqnarray}  
    M_{zz}(n, t)
    &
    = & 
    \frac{1}{t}
    \frac{1}{n^2}
    \sum_{\ell_1=1}^n \sum_{\ell_2=1}^n
    \int_{0}^{t} \mathrm{d} t'
    \langle \psi(t') \vert
    \sigma_{\ell_2}^{z} \sigma_{\ell_1}^{z}
    \vert \psi(t') \rangle \nonumber
    \\  &
    \equiv &
    \frac{1}{t}
    \frac{1}{n^2}
    \int_{0}^{t} \mathrm{d} t' \langle 4S_{z}^{2} \rangle (t')
    ,
  \label{eq:MzzOrderParameter}
\end{eqnarray}
which in the thermodynamic limit and long-time limit exhibits a minimum and discontinuity in the slope as a function of $B/J_0$ at the phase transition. At finite system size $n$, the discontinuity in the slope is smoothed out~\cite{zhang2017observationmany, de2023non}. The minimum of $M_{zz}$ with respect to the quenched transverse field $B$ is a robust signature of the DQPT and corresponds to the critical point of the phase transition in the thermodynamic liit $n \to \infty$. We elaborate on this using a semiclassical picture of the fully connected $\alpha = 0$ model in Appendix~\ref{app:DQPTSimplePicture}. The minimum tends to be less sensitive to taking the long-time limit than the phase transition in $M_z(n,t)$, hence it is a more convenient observable to use to identify the critical parameters.

We focus on the case of $\alpha = 1.5$ because it is phenomenologically far from both the collective-spin regime ($\alpha = 0$) and one-dimensional nearest neighbor model ($\alpha \to \infty$). We use time evolution of MPS with the time dependent variational principle (TDVP)~\cite{haegeman2011time, haegeman2016unifying} to calculate the time evolution of the quantum state and the observables $M_z(n,t)$, Eq.~\eqref{eq:MzOrderParameter}, and $M_{zz}(n,t)$, Eq.~\eqref{eq:MzzOrderParameter}, in order to extract the critical properties in the thermodynamic limit and long-time limit. We provide details of the simulations in Appendix~\ref{app:TDVP}. We will show that estimating these quantities is significantly easier classically, by simulating them with highly truncated MPS. We shall see that the robustness of the order parameters to bond dimension truncation arises because they involve spatial and temporal averaging and exhibit fast equilibration, which allows for accurate estimates at relatively short times. 

\begin{figure}[htbp] 
   \centering
    \subfigure[]{\includegraphics[width=0.48\textwidth]{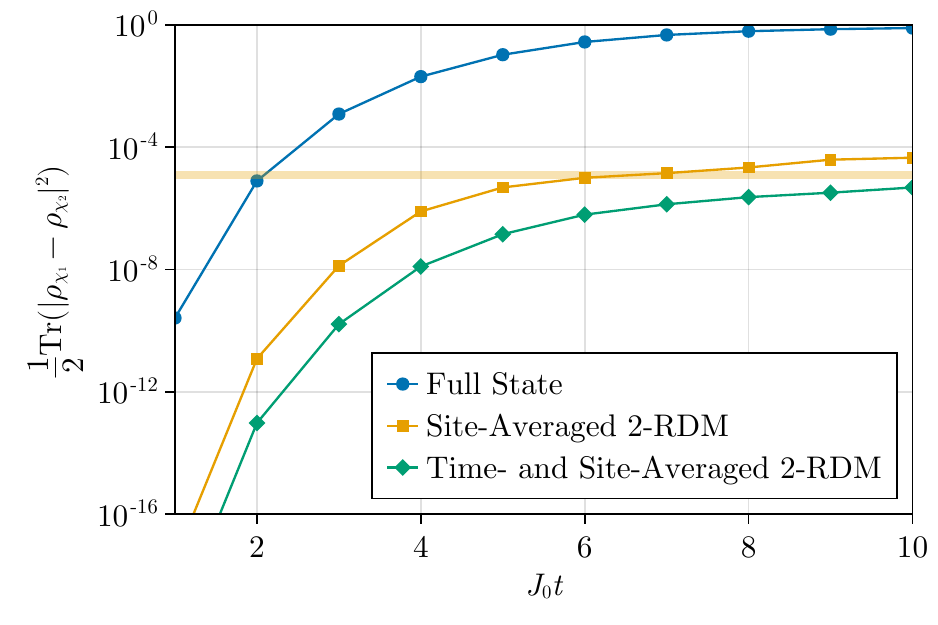}\label{fig:HSD}}
   \subfigure[]{\includegraphics[width=0.48\textwidth]{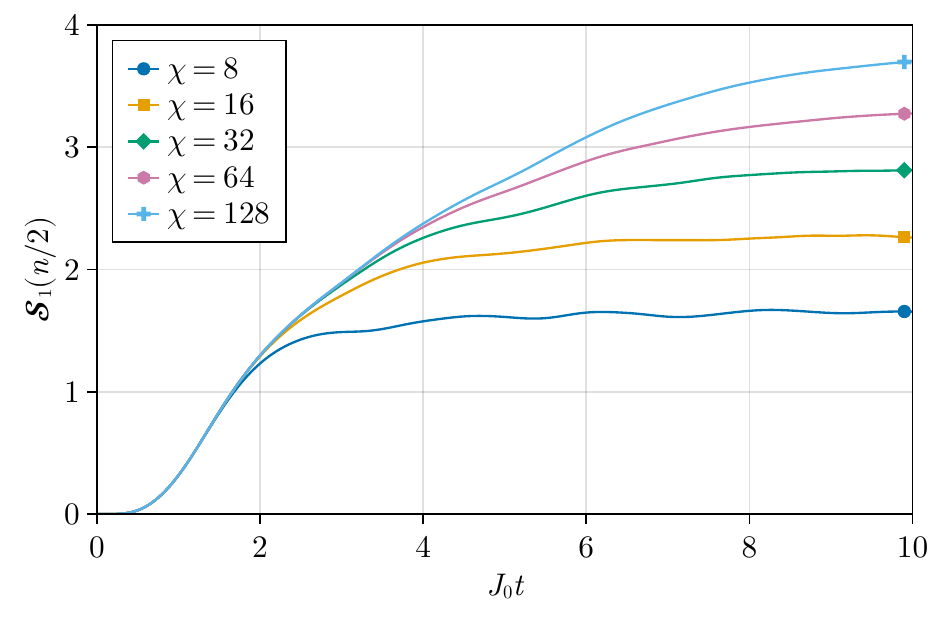}\label{fig:EE}}
   \caption{{The role of entanglement and bond-dimension truncation in approximating the quantum state.} (a) The squared HS distance, Eq. (\ref{Eq:HS-Distance}), for 2-site reduced density operator for $\chi_1 = 64$, and $\chi_2 = 128$. Other simulation parameters $n=50, \alpha=1.5, B/J_0 = 1, J_0 \Delta t = 10^{-2}$. The band corresponds to the site-averaged result for random MPS, sampled from the Gaussian tensor ensemble. The time- and site-averaged reduced density operator has an almost 6 orders of magnitude smaller squared HS distance squared compared to the full state. (b) Half-system entanglement entropy, $\mathcal{S}_1$, for different bond dimension simulations. When a curve deviates from the rest gives an estimate of when the associated bond dimension is no longer sufficient to track the entanglement of the many-body state.}
   \label{fig:AbsoluteError}
\end{figure}

We first contrast approximating the full state (the microstate) to approximating site- and time-averaged reduced density matrices, which determine the macroproperties of $M_{z}$ and $M_{zz}$. To do this, we consider the squared overlap between the states $\ket{\psi_{\chi_1}(t)}$ and $\ket{\psi_{\chi_2}(t)}$ generated by our MPS simulations with maximum bond dimension $\chi_1$ and $\chi_2$ respectively,
\begin{equation}
    1 - \left| \braket{\psi_{\chi_1}(t)}{\psi_{\chi_2}(t)} \right|^2 = \frac{1}{2}\mathcal{D}^2_{\mathrm{HS}}\big(\rho_{\chi_1}(t),\rho_{\chi_2}(t)\big),
    \label{Eq:HS-Distance}
\end{equation}
where we have related the squared HS distance (Eq.~\eqref{eq:HSDistanceDefinition}) to the squared overlap of the full many-body pure state. We show this infidelity in Fig.~\ref{fig:HSD} between bond dimensions $\chi = 64$ and $\chi = 128$ for a system of $n=50$ spins, where for $J_0 t \approx 5$ we see that the infidelity is above $10^{-1}$. Moreover, we see in Fig.~\ref{fig:EE} a difference in the half-system entanglement entropy $\mathcal{S}_1$ of the states evaluated with different bond-dimensions, as $\mathcal{S}_1 \leq \mathcal{S}_0 = \ln(\chi)$, showing that the fixed bond dimension simulations for $\chi \leq 64$ are clearly insufficient for capturing the full entanglement structure of the full quantum state for $J_0 t \gtrsim 5$.

We contrast this with the squared HS distance of the 2-site reduced density operators associated with the states $\ket{\psi_{\chi_1}(t)}$ and $\ket{\psi_{\chi_2}(t)}$ after the spatial and time averaging. The HS distance serves as an upper bound for the error of observables, as discussed in Appendix~\ref{app:LocalReducedStates}. In addition to the squared HS distance between the full states with $\chi = 64$ and $\chi = 128$, we show in Fig.~\ref{fig:HSD} the squared HS distance for two additional different cases:
\begin{enumerate}
    \item first averaging all instantaneous two-spin reduced density operators (Site-Averaged 2-RDM), then calculating the squared HS distance;
    \item first calculating the time-averaged two-spin reduced density operator's, then averaging over sites  (Time- and Site-Averaged 2-RDM), and then calculating the squared HS distance.
\end{enumerate}

Spatial and temporal averaging significantly reduce the error, and at $J_0 t \approx 5$, the squared HS distance is closer to $10^{-7}$ for the 2-site averaged reduced density operator. This indicates that spatial and temporal averaging allow for more accurate estimates of order parameters for this system than the full state infidelity might suggest. We also consider a randomly generated MPS, using the Gaussian tensor ensemble (see Appendix~\ref{app:RandomMPS} for details), and note that randomly generated MPS with $n=50$ sites and with bond dimensions $64$ and $128$ have (on average) a space-averaged squared HS distance of $10^{-5}$.

\begin{figure}[ht] % 
   \centering
   \includegraphics[width=0.48\textwidth]{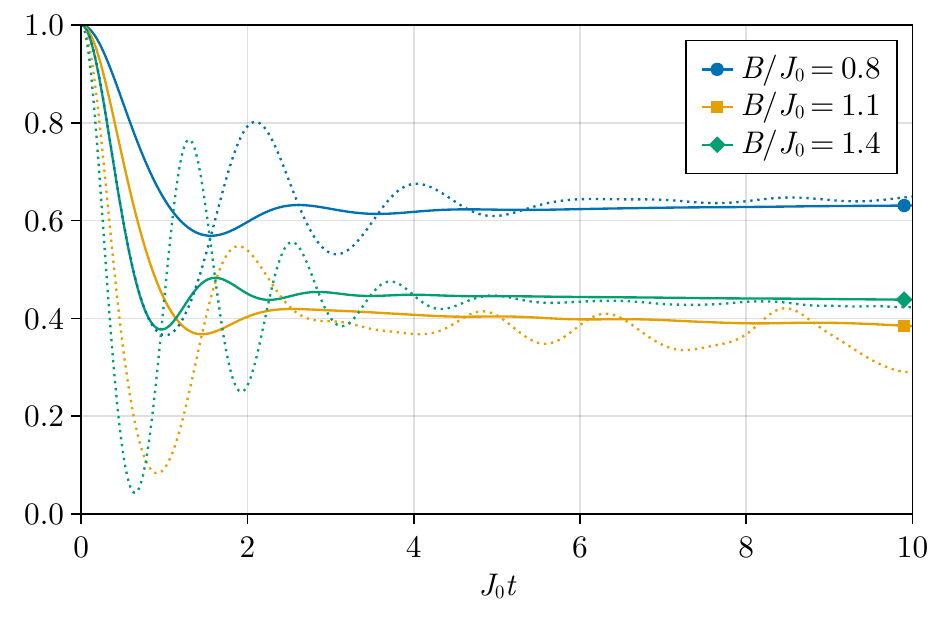} 
   \caption{The evolution of the time-averaged magnetization squared $M_{zz}(n,t)$ (thick lines) and the instantaneous magnetization squared $\langle 4S_z^2 \rangle(t)/n^2$ (dotted lines). Simulation parameters are $n=50, \alpha = 1.5, J_0 \Delta t = 10^{-2}, \chi = 128$. The time-averaged observable is already close to its steady-state value by $J_0 t = 3$.}
   \label{fig:Mzz}
\end{figure}

Another crucial ingredient for the classical tractability of the critical properties of the DQPT is the relatively rapid equilibration of $M_{zz}$ (we elaborate more on this property in Sec.~\ref{sec:RoleChaos}), allowing us to estimate long-time-averaged expectation values from relatively short time simulations. To illustrate this property, we visualize in Fig.~\ref{fig:Mzz} the behavior of both $\langle 4 S_{z}^2 \rangle(t)$ and $M_{zz}(t)$ for different quench parameters $B/J_0$ starting from the initial state $\ket{\uparrow_z}$. After $J_0 t \approx 3$, the instantaneous observable oscillates close to a fixed value and the time-average approaches this value steadily. As we will demonstrate next, this allows for the extraction of critical properties of the DQPT using simulations with $3 \leq J_0 t \leq 10$ and with surprisingly low bond dimensions. 
%

%
%%%%%%%%%%%%%%%%%%%%%%%%%%%%%%%%%%%%%%%%%%%%%%%%%%%%%%%%%%%%%%%%%%%%%%%%%%%%%
\subsection{Estimating the critical point of a DQPT using the minimum of $M_{zz}(t)$} \label{sec:DQPTmin}
\begin{figure}[htbp] % 
   \centering
   \subfigure[]{\includegraphics[width=0.48\textwidth]{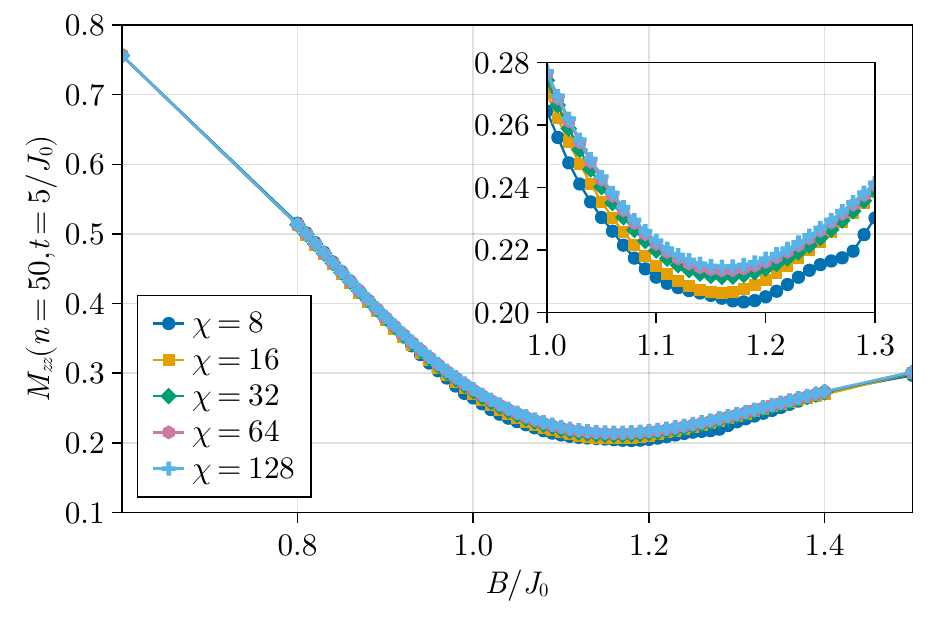} \label{fig:DQPTParabola1}}
   \subfigure[]{\includegraphics[width=0.48\textwidth]{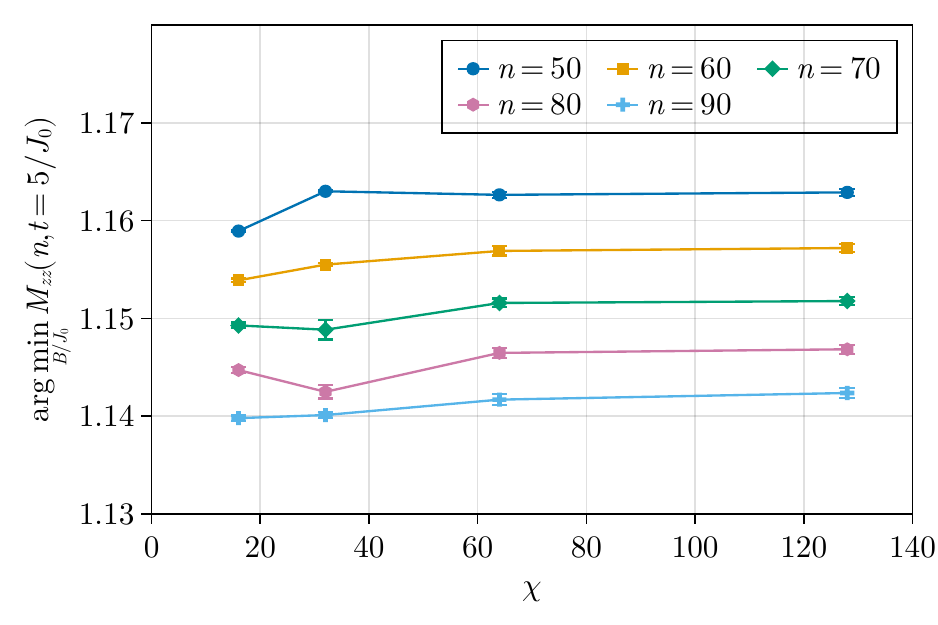} \label{fig:DQPTParabolaFit1}} 
   \caption{Estimating the critical point using the location of the minimum of $M_{zz}$ as a function of $B/J_0$.  (a) $M_{zz}$ versus $B/J_0$ for $n=50$, $\alpha = 1.5$, $J_0 t = 5$, and different bond dimension $\chi$. (b) Fits of the location of the minimum of $M_{zz}$ with respect to $B/J_0$ for $\alpha = 1.5$, $J_0 t = 5$, different $n$ and bond dimension $\chi$; error bars are smaller than the marker size. These plots show the robustness to MPS truncation in the estimation of the minimum in the order parameter, even for a relatively large number of spins.}
   \label{fig:DQPTmin}
\end{figure}

%%%%%%%%%%%%%%%%%%%%%%%%%%%%%%%%%%%%%%%%%%%%%%%%%%%%%%%%%%%%%%%%%%%%%%%%%%%%%
%
As a first demonstration of the robustness of estimating critical properties, we estimate the critical transverse field magnitude $(B/J_0)_{\mathrm{c}}$ by identifying the minimum in the curve of $M_{zz}$ as a function of $B/J_0$. We see the behavior of $M_{zz}(n,t)$ in Fig.~\ref{fig:DQPTParabola1} at a final quench time of $J_0t = 5$ for $n=50$ for different bond dimensions. By fitting the points around the minimum with a parabola, we can extract the minima as a function of $n$ and $\chi$, an shown in Fig.~\ref{fig:DQPTParabolaFit1}. Importantly, we observe that the predicted $(B/J_0)_{\mathrm{c}}$ has a weak dependence on the bond dimension for $\chi \geq 16$ for a fixed system size $n$. Additionally, we see an almost linear behavior as a function of system size for a fixed bond dimension, indicating that we are still seeing some finite size effects for $n\leq 90$. We emphasize even though the overlap between the MPS with bond dimensions $\chi$ and $2 \chi$ is very small, they still give very similar estimates for the location of the minimum, and thus the critical point. As an example, for $n=90$ the overlap squared between states with bond dimension $\chi =16$ and $\chi = 32$ is $< 10^{-6}$ (not shown) but their estimates of the location of the minimum are within $10^{-2}$ of each other.

In Appendix~\ref{app:DQPTCriticalPointMinimum}, we repeat the analysis for a later time $J_0 t = 10$, and we observe that changing the final time $J_0 t$ changes the overall value of $M_{zz}(n,t)$ but does not significantly change the location of the minimum of $M_{zz}$ for bond dimensions $\chi \geq 16$. Thus our results suggest that a bond dimension of $\chi = 16$ and a quench time of $J_0 t = 5$ should be sufficient for identifying the critical point via the minimum of $M_{zz}$ at arbitrary system sizes. 

\begin{figure}[htbp] % 
   \centering
   \includegraphics[width=0.46\textwidth]{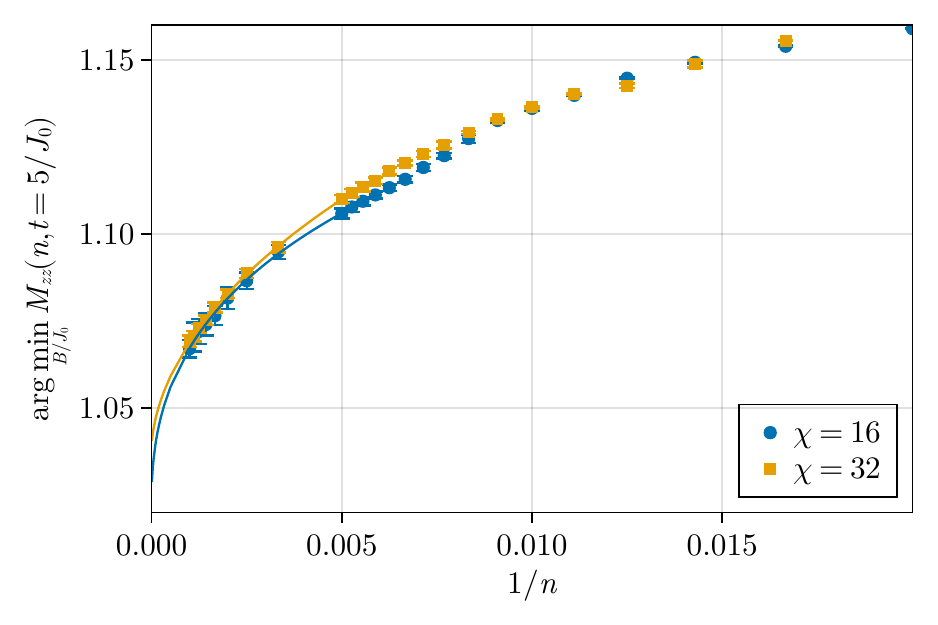} 
   \caption{Fits of the the location of the minimum of $M_{zz}$ with respect to $B/J_0$ for $\alpha = 1.5$, $J_0 t = 5$, different $n$ and bond dimensions, $\chi =16$ and $32$. The solid lines correspond to a fit to the function $a + b n^{-c}\ln(n)$ using $n \geq 200$, with $ a = 1.02 \pm 0.04, b = 0.3 \pm 0.3, c = 0.5 \pm 0.3$ for $\chi = 16$ and $ a = 1.04 \pm 0.02, b = 0.5 \pm 0.4, c = 0.7 \pm 0.2$ for $\chi = 32$. The estimates for the location of the minimum in the thermodynamic limit are within the uncertainty of the fits.}
   \label{fig:DQPTminLargeFixedBonddim}
\end{figure}

Using the above hypothesis, in Fig.~\ref{fig:DQPTminLargeFixedBonddim} we show our results for the minimum of $M_{zz}$ up to $n=1000$ spins using $\chi = 16$ and $\chi = 32$, where the data at large sizes allows us to extrapolate the location of the critical point in the thermodynamic limit. 


The weak dependence of the minimum of \(M_{zz}\) with bond dimension observed in Fig.~\ref{fig:DQPTmin}
%with respect to the quench ratio \(B/J_0\) 
can be explained by understanding how the order parameter \(M_{zz}\) is influenced by space and time-averaged expectations of local observables. The process of averaging over space and time serves to smooth out fluctuations in individual two-spin expectation values, rendering them less susceptible to a precise specification of the quenched microstate. As a result, the expectation values become less prone to imperfect specification of the microstate, allowing for more accurate determinations of physical quantities of interest using heavily truncated MPS and noisy experiments. This effect can be likened to the self-averaging behavior observed in individual spins within a large spin system with many interacting neighbors, where the influence of neighboring spins leads to a suppression of noise and variability.
%. In such systems, neighboring spins exert an influence that leads to a suppression of noise and variability. Similarly, when considering the averaging process for two-spin expectations, the influence of local fluctuations is diminished due to the smoothing effect of space and time averaging. 

This phenomenon has significant implications for our understanding of spin systems, their response to external influences, and the tractability of critical phenomena using classical approximate methods. By examining the effects of space and time averaging on local observables, we can gain insights into the behavior of individual or a small number of spins within a larger system, ultimately shedding light on the underlying mechanisms driving the system's dynamics. In Sec.~\ref{sec:CorrelationLengthFromQuench}, we consider a scenario where the quantities of interest involve two-spin correlations along the $z$-direction, like \(M_{zz}\), but without space and time averaging. We will see that this leads to more sensitivity to the microstate.

%%%%%%%%%%%%%%%%%%%%%%%%%%%%%%%%%%%%%%%%%%%%%%%%%%%%%%%%%%%%%%%%%%%%%%%%%%%%%
\subsection{Finite scaling analysis and universality}
\label{subsec:FiniteSizeScaling}
%%%%%%%%%%%%%%%%%%%%%%%%%%%%%%%%%%%%%%%%%%%%%%%%%%%%%%%%%%%%%%%%%%%%%%%%%%%%%
%
To further study the critical behavior near the DQPT in the thermodynamic limit, we perform a finite-size scaling (FSS) analysis. To do so we use the data collapse method \cite{New1999,Bin2010} implemented in the \emph{pyfssa} scientific software package \cite{And2015,Mel2009} on the order parameter $M_z$, and we consider two different FSS ansatz.

The first is an ``equilibrium'' ansatz, where we assume that the behavior of $M_{z}$ and $M_{zz}$ near the DQPT for sufficiently long quench times $J_0 t$ only depends on the system size $n$ and not on the quench time. While this seems reasonable given our observation from the previous section that $M_{zz}$ thermalizes rapidly and exhibits a weak dependence on the quench time $J_0 t$, the long-time limit of the FSS ansatz assumes global equilibration and not only local equilibration, which is unlikely to be satisfied for our relatively short time simulations of $J_0 t \ll n$. We will focus on the case of $M_{z}$ as it exhibits more robust predictions as a function of $J_0 t$ for the critical parameters, suggesting less sensitivity to this equilibrium assumption. We discuss the FSS analysis for $M_{zz}$ in Appendix~\ref{app:FiniteSizeScalingMz}.

\begin{figure*}[htbp] % 
   \centering
   \subfigure[]
   {\includegraphics[width=0.48\textwidth]{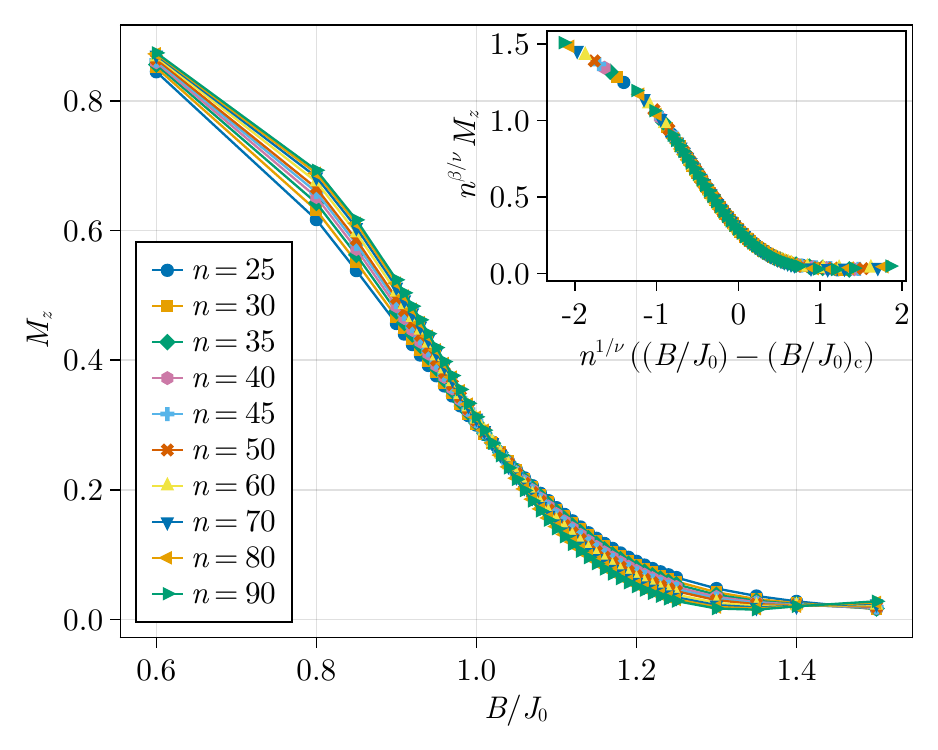} \label{fig:DataCollapseMzSpaceExample}}
   \subfigure[]
   {\includegraphics[width=0.48\textwidth]{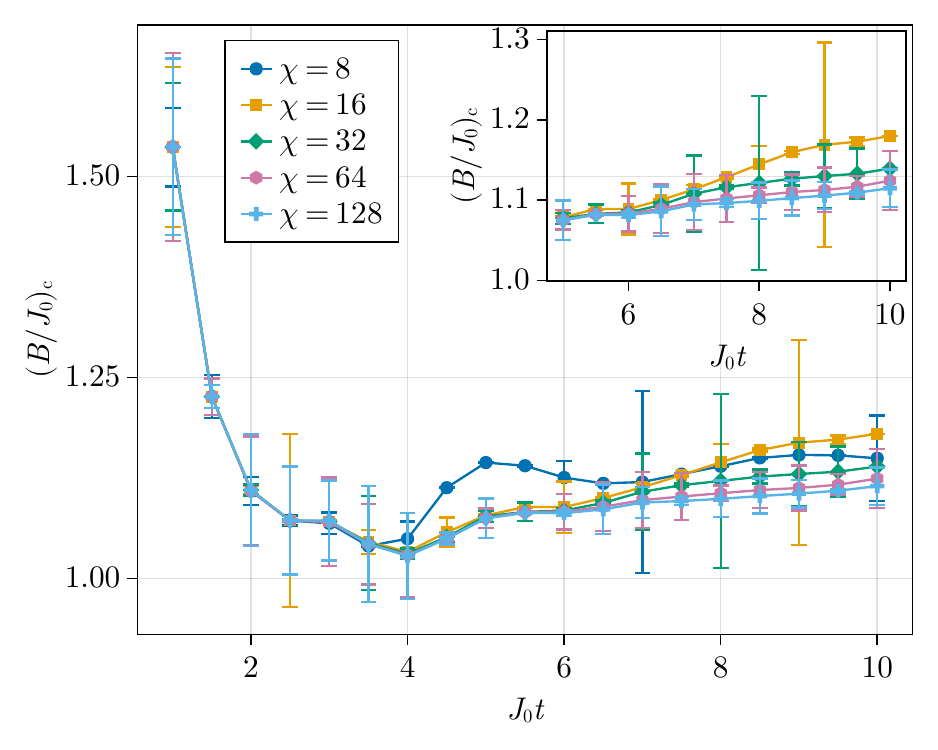} \label{fig:DataCollapseSpaceMzBCritical}}
   \subfigure[]
   {\includegraphics[width=0.48\textwidth]{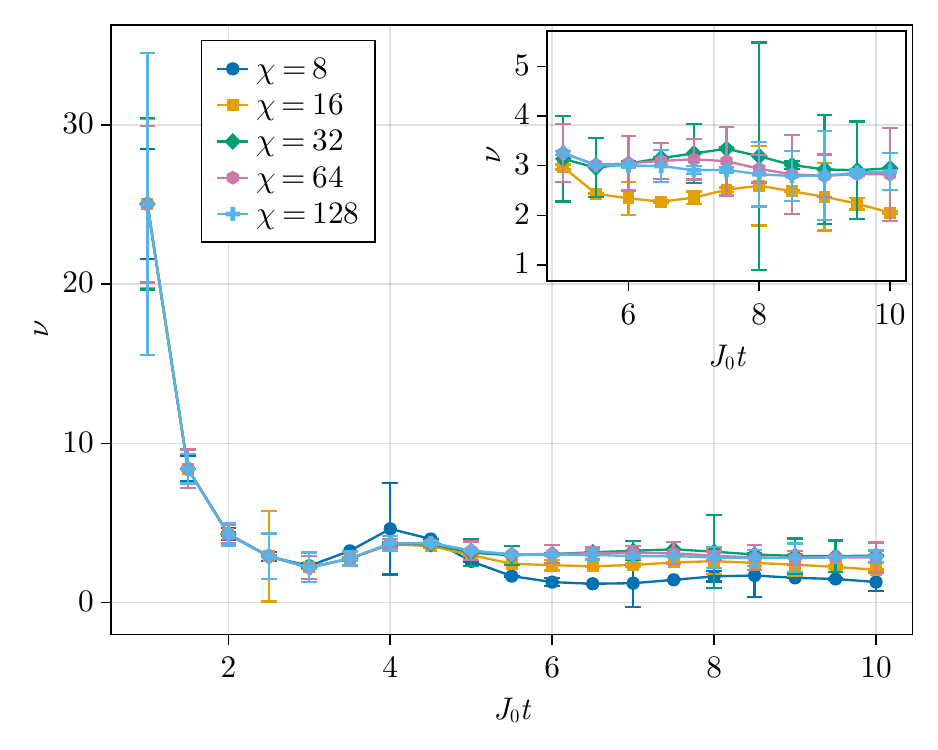} \label{fig:DataCollapseMzSpaceNu}}
   \subfigure[]
   {\includegraphics[width=0.48\textwidth]{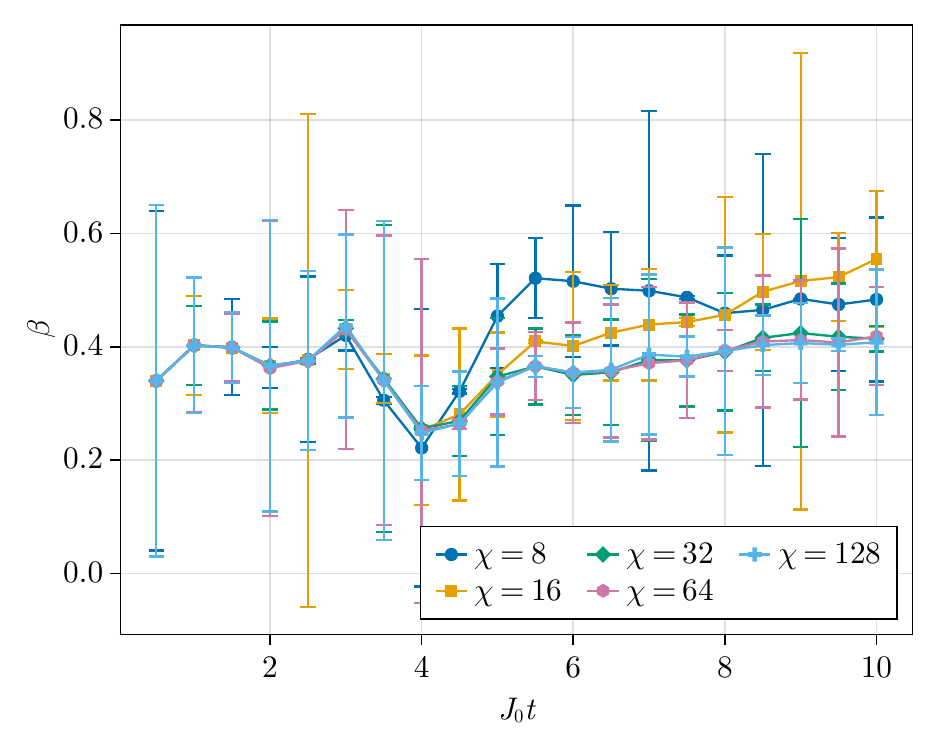} \label{fig:DataCollapseMzSpaceZeta}}
   \caption{Estimates of critical properties using the finite system size scaling ansatz [Eq.\eqref{eq:FiniteSizeScalingSpaceAnsatz}]. (a) Finite-size scaling analysis for the order parameter $M_{z}$ for simulations with parameters $\alpha = 1.5, J_0 t = 5.5, \chi = 64, J_0 \Delta t = 10^{-2}$. Using the data collapse method, the extracted critical parameters from this simulation for this total simulation time are $\left(B/J_0\right)_{\mathrm{c}} = 1.082 \pm 0.001, \nu = 3.008 \pm 0.008, \beta = 0.37 \pm 0.02$. We further study the dependence of the critical point (b) and critical exponents (c),(d),  on the total quench time for different bond dimensions $\chi \in \{8, 16, 32, 64, 128\}$. Insets show estimates for $J_0t \geq 5.0$, which are more consistent. Though the order parameter is defined in the asymptotic limit, it converges at relatively short times. The estimates of the critical point and critical exponents also converge at relatively small bond dimensions.} 
   \label{fig:DataCollapseParametersMzSpace}
\end{figure*}

We take as our equilibrium FSS ansatz
\begin{equation}
    M_{z}(n) = n^{-\beta/\nu} f\left(n^{1/\nu} \,
    \left(\frac{B}{J_0} - \left(\frac{B}{J_0}\right)_{\mathrm{c}}\right); J_0 t \right),
    \label{eq:FiniteSizeScalingSpaceAnsatz}
\end{equation}
where $f$ is a universal scaling function and the exponents $\beta$ and $\nu$ are the critical exponents that characterize the behavior of $M_{z}$ near the DQPT. Specifically, near the critical point and in the thermodynamic limit, the exponent $\nu$ characterizes the scaling of correlation lengths $\xi \sim | B/J_0 - (B/J_0)_{\mathrm{c}}|^{-\nu}$, and $\beta$ characterizes the scaling of $M_{z}(\infty)$ near the critical point:
\begin{equation}\label{eq:fssuniversal}
 f(x) \sim \begin{cases}
     |x|^{\beta} & \text{for }  x \to -\infty , \\
     \mathrm{const.} & \text{for } x \to +\infty .
 \end{cases}
\end{equation}
The data collapse method proceeds by identifying the critical parameters $\left\{ (B/J_0)_{\mathrm{c}}, \beta, \nu\right\}$ that give universal behavior for $M_{z}(n) n^{\beta/\nu}$ as a function of $n^{1/\nu} \left( B/J_0 - (B/J_0)_{\mathrm{c}} \right)$. We show an example of the results of the data collapse for a fixed $J_0 t$ in Fig.~\ref{fig:DataCollapseMzSpaceExample}.

We perform the data collapse and identify the critical parameters for different fixed $J_0 t$ and bond-dimension $\chi$, which we show in Figs.~\ref{fig:DataCollapseSpaceMzBCritical}-\ref{fig:DataCollapseMzSpaceZeta}. For short duration quenches, $J_0t \lesssim 2$, the FSS results are inconsistent with those for longer quenches, implying that such time scales are too short to be used for FSS.
For longer duration quenches $J_0t \gtrsim 5$, the estimates of the critical parameters are more consistent with each other, exhibiting a negligible drift with $J_0 t$ and showing agreement within the error bars for bond dimensions $\chi \geq 16$ at a fixed $J_0 t$. (We note that the critical point $\left(B/J_0 \right)_{\mathrm{c}}$ is already well estimated for simulation times of $J_0 t \gtrsim 3$.) The estimate of $(B/J_0)_{\mathrm{c}} \approx 1.1$ is consistent with the analysis of the minimum of $M_{zz}$ as a function of $B/J_0$ performed in the previous section. This shows that relatively rapid equilibration allows us to estimate the critical point and critical exponents at short times $J_0 t \ll n$. 
Furthermore, since the estimates of the critical parameters using $\chi = 16$ agree to within the error bars with the estimates using $\chi = 128$, it suggests that many-body entanglement beyond a R{\'e}nyi-0 entropy of $\mathcal{S}_0 = \ln(16)$ plays a negligible role in this phase transition. 

\begin{figure*}[htbp] % 
   \centering
   \subfigure[]
              {\includegraphics[width=0.48\textwidth]{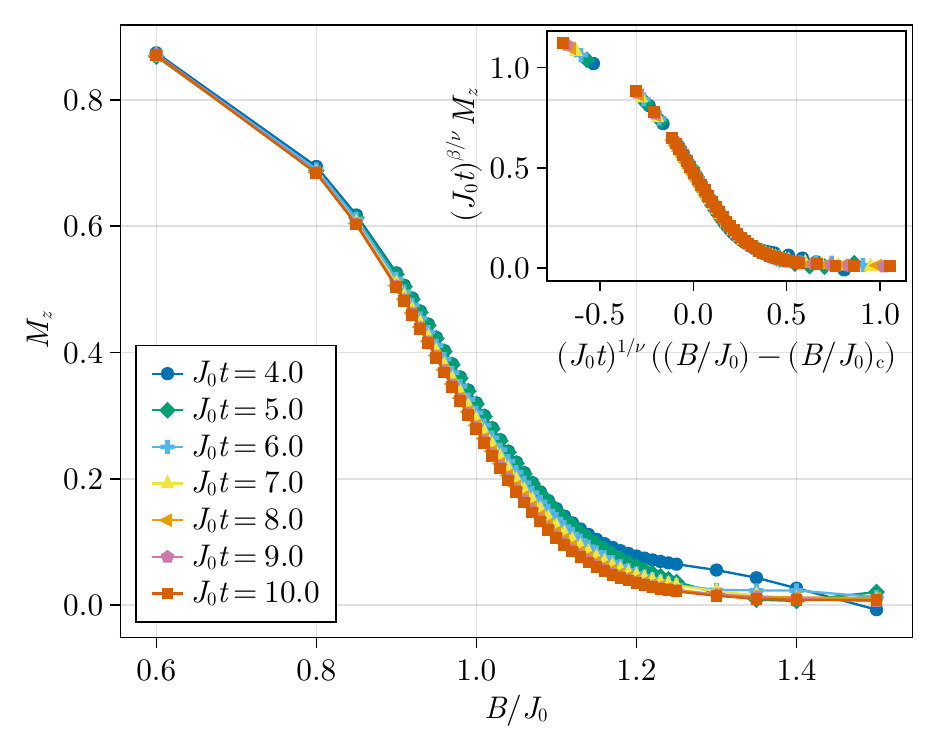} \label{fig:DataCollapseMzTimeExample}}
   \subfigure[]
             {\includegraphics[width=0.48\textwidth]{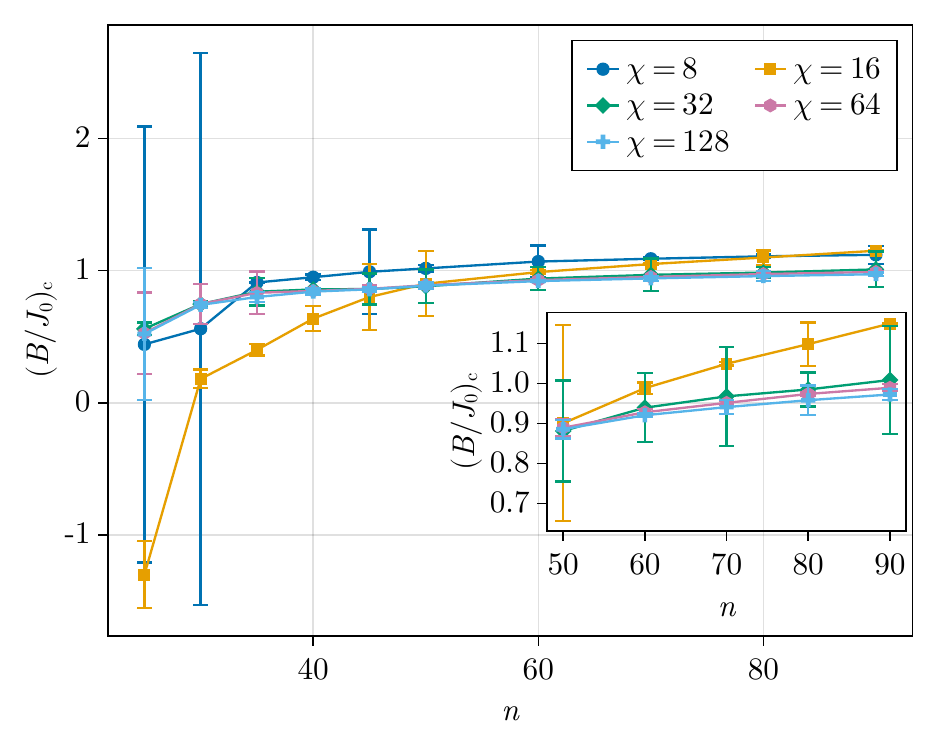} \label{fig:DataCollapseMzTimeBCritical}}
   \subfigure[]
             {\includegraphics[width=0.48\textwidth]{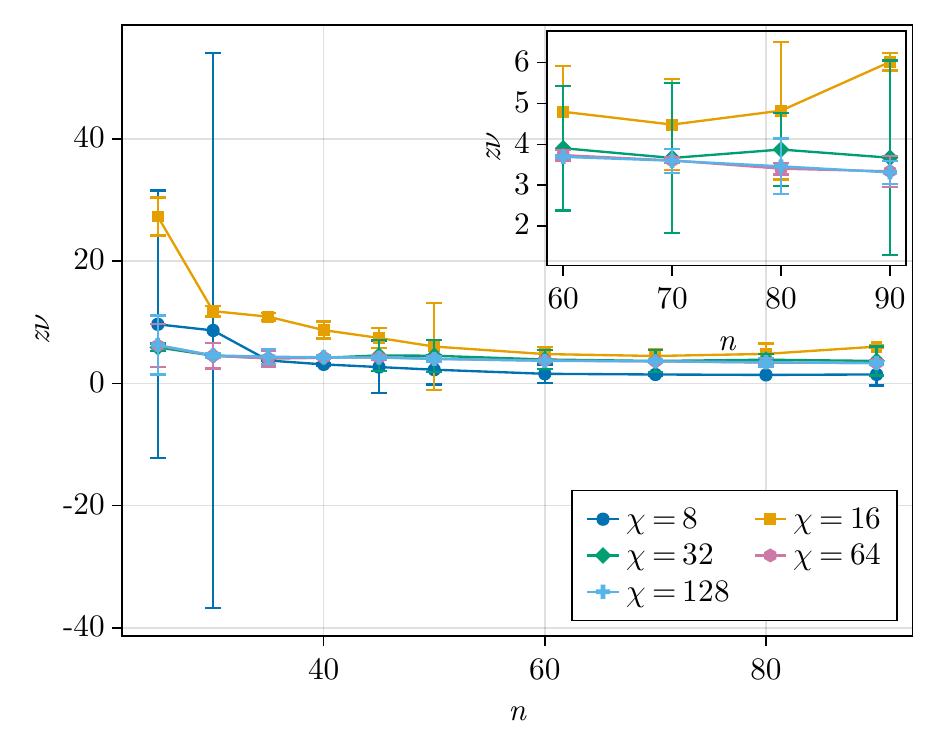} \label{fig:DataCollapseMzTimeNu}}
   \subfigure[]
             {\includegraphics[width=0.48\textwidth]{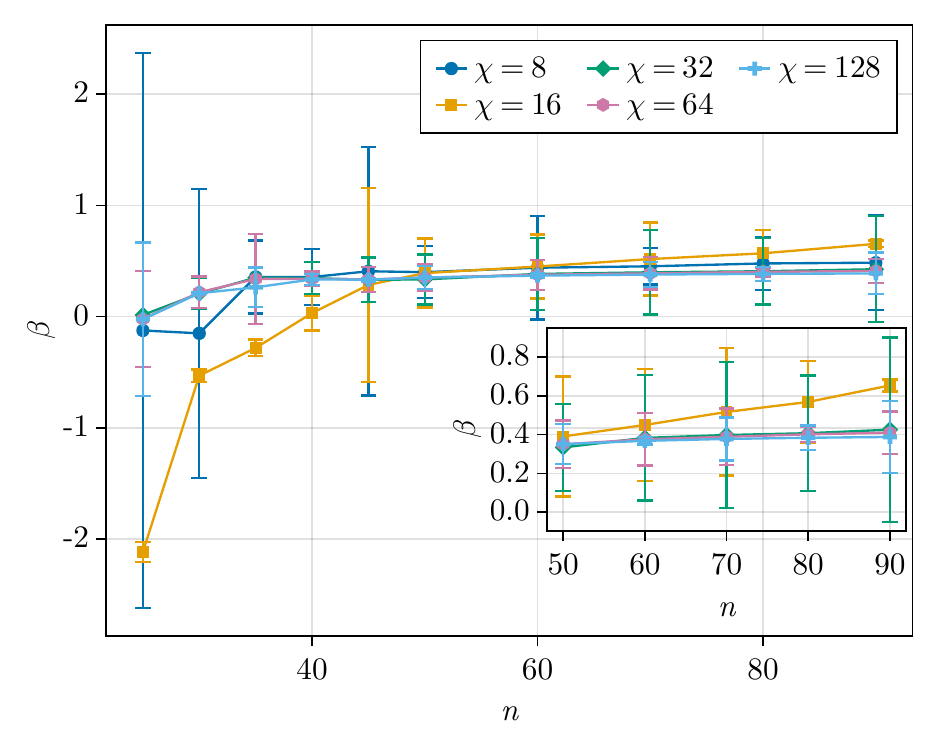} \label{fig:DataCollapseMzTimeZeta}}
    \caption{Estimates of critical properties using a finite quench duration scaling ansatz [Eq.\eqref{eq:FiniteSizeScalingSpaceAnsatz}]. (a) Finite-size scaling analysis for order parameter $M_{z}$ for simulations using
	$\alpha =1.5, n = 80, \chi = 128, \Delta t = 10^{-2}$. Using the data collapse method, the extracted critical parameters for these simulations parameters from this simulation are $\left(B/J_0\right)_{\mathrm{c}} = 0.95 \pm 0.04, z\nu = 3.5 \pm 0.7, \beta = 0.38 \pm 0.6$. Using the data collapse method, the extracted critical point (b) and critical exponents (c), (d) as a function of system size $n$, for different bond dimensions $\chi \in \{8, 16, 32, 64, 128\}$. Insets show estimates for $n \geq 50$, which are more consistent. The estimates of the critical point and critical exponents are approximately the same across different bond-dimensions.}
   \label{fig:DataCollapseParametersMzTime}
\end{figure*}

For our second FSS analysis, we choose a ``nonequilibrium'' FSS ansatz that reflects the fact that our simulations are relatively short time simulations with $J_0 t \ll n$. Nonequilibrium data for FSS is known to be possible in a Monte-Carlo setting \cite{ozeki2007nonequilibrium}. We consider a nonequilibrium FSS ansatz with no system size dependence (we have now effectively taken the infinite size limit)
\begin{equation}
    M_{z}(t) = (J_0 t)^{-\beta/(z\nu)} \,
    g\left((J_0 t)^{1/(z\nu)} \left(\frac{B}{J_0} - \left(\frac{B}{J_0}\right)_{\mathrm{c}}\right);
	n \right),
    \label{eq:FiniteSizeScalingTimeAnsatz}
\end{equation}
where $g$ is a universal scaling function. Similar to our first FSS ansatz (Eq.~\eqref{eq:FiniteSizeScalingSpaceAnsatz}), near the critical point and in the infinite time limit we expect a diverging time scale $\tau \sim | B/J_0 - (B/J_0)_{\mathrm{c}}|^{-z \nu}$ with dynamical critical exponent $z \nu$. (We note that a critical slowing down is observed for the $\alpha = 0$ case as we approach the critical point.) We expect $g$ to behave in a similar manner to $f$, that is,
\begin{equation}
 g(x) \sim 
    \begin{cases}
        |x|^{\beta} & \text{for }  x \to -\infty , \\
        \mathrm{constant} & \text{for } x \to +\infty .
    \end{cases}
\end{equation}
The data collapse method proceeds similar to before by identifying the critical parameters that give universal behavior for $M_{z}(t) (J_0 t)^{\beta/(z \nu)}$ as a function of $(J_0 t)^{1/(z\nu)} \left( B/J_0 - (B/J_0)_{\mathrm{c}} \right)$. We show an example of the results of the data collapse method in Fig.~\ref{fig:DataCollapseMzTimeExample}.

We perform the data collapse and identify the critical parameters for different system sizes $n$ and bond-dimension $\chi$, which we show in Figs.~\ref{fig:DataCollapseMzTimeBCritical}-\ref{fig:DataCollapseMzTimeZeta}. The data collapse is not as successful as for the equilibrium ansatz, suggesting that we are not as close to the steady state limit $J_0 t \to \infty$ as we are to the thermodynamic limit $n \to \infty$. For small system sizes $n < 50$ the FSS results are inconsistent with those for larger system sizes, suggesting that the assumption that $J_0 t \ll n$ is being violated for these smaller system sizes. For larger sizes $n \gtrsim 50$ the estimates of the critical parameters are more consistent with each other, exhibiting a negligible drift with $n$ and generally showing agreement within the error bars for bond dimensions $\chi \geq 32$ at a fixed $n$.

Our simulations show that the order parameters corresponding to the DQPT can be approximated classically using modest bond-dimension MPS ($\chi \ll 2^{n/2}$), with a runtime that scale as $\mathcal{O}(n \chi^3)$. The simulations allow us to not only distinguish the symmetry-broken phase from the symmetry-restored phase but to also estimate the critical transverse field and critical exponents of the phase transition. The short duration of the quenches considered in our simulations affect our analysis of both order parameters, $M_{z}$ and $M_{zz}$. The drift in critical exponents for $M_{zz}$ as a function of $J_0 t$ is more pronounced, and the finite quench duration scaling for $M_{z}$ is not as successful as the finite size scaling. While a careful analysis of how these estimated values converge as a function of the computational cost is beyond the scope of this work, we believe that these methods can be used to quantitatively study properties of the phenomenon without the overhead of an exponentially large state space. A detailed understanding of the critical behavior near the DQPT across different values of $\alpha$ remains an avenue of future work.

%

%
%%%%%%%%%%%%%%%%%%%%%%%%%%%%%%%%%%%%%%%%%%%%%%%%%%%%%%%%%%%%%%%%%%%%%%%%%%
\section{Estimation of correlation lengths from short-time quench dynamics}
\label{sec:CorrelationLengthFromQuench}
%%%%%%%%%%%%%%%%%%%%%%%%%%%%%%%%%%%%%%%%%%%%%%%%%%%%%%%%%%%%%%%%%%%%%%%%%%
%
As a second example, we consider a problem involving two-point correlations, but without space- and time averaging as for the case of DQPT. In particular, we simulate quench dynamics to extract the long-time correlation length in the integrable 1D nearest-neighbor TFIM
 \begin{equation} \label{eqt:alphaInftyTFIM}
  H =
  - J_0 \sum_{\ell=1}^{n-1} \sigma_\ell^z \sigma_{\ell+1}^z
  - B \sum_{\ell=1}^n \sigma_\ell^x,
\end{equation}
which corresponds to taking $\alpha \rightarrow \infty$ in Eq.~\eqref{eq:HamiltonianPowerLawTFIM}. The local expectation value of interest is the correlation function 
\begin{equation} \label{eqt:Czz}
  C_{zz}(t, \ell) = \bra{\psi(t)}\sigma_b^z \sigma_{b + \ell}^z  \ket{\psi(t)},
\end{equation}
where $b$ is the index of the center of the chain with open boundary conditions (to avoid ambiguity, we consider chains of odd length). In the thermodynamic limit $n \rightarrow \infty$, and in the infinite-time limit when correlations have completely spread through the chain and equilibrated, when quenching from initial field $B_{\mathrm{i}}$ to final field $B_{\mathrm{f}}$ the correlation function takes the form~\cite{calabrese2012quantum1, calabrese2012quantum2}
\begin{equation}
    C_{zz}(t_\infty, \ell) \approx
    C_0(B_{\mathrm{f}},B_{\mathrm{i}}) 
    \exp\left(-\frac{\ell}{\xi(B_{\mathrm{f}},B_{\mathrm{i}})}\right) ,
\end{equation}
where $C_0(B_{\mathrm{f}},B_{\mathrm{i}})$ is given by
\begin{equation}
C_0(B_{\mathrm{f}},B_{\mathrm{i}}) = \left[ \frac{(B_{\mathrm{i}}- B_{\mathrm{f}}) B_{\mathrm{f}} \sqrt{B_{\mathrm{i}}^2-1}}{(B_{\mathrm{i}}+ B_{\mathrm{f}})(B_{\mathrm{i}}B_{\mathrm{f}}-1)}\right]^{1/2} .
\end{equation}
Of particular interest is the correlation length $\xi$ at the critical point $B_{\mathrm{f}} = J_0$ (for $\alpha \to \infty$) of the ground state quantum phase transition. Near the critical point we expect the quench dynamics to be the hardest to simulate as entanglement grows most strongly. Moreover, this point is of most interest as it can be related to its thermal counterpart, $\xi_\mathrm{T}$, and used to extract the universality of the transition exhibited in the model~\cite{Kar2017, Tit2019}. Because this problem is integrable, long-studied known solutions~\cite{derzhko1998numerical, calabrese2011quantum, calabrese2012quantum1, calabrese2012quantum2} can be used to benchmark quantum simulators. 

While the exact correlation length is defined in the infinite-time limit, Karl {\em et al.} showed that, surprisingly, $\xi(B_{\mathrm{f}},B_{\mathrm{i}})$ can be extracted from {\em short-time quenches} due to the different time scales for local and global equilibration~\cite{Kar2017}. This gives us an opportunity to explore the robustness to truncation in the MPS simulation under consideration here. Moreover, our previous results for the $\mathbb{Z}_2$ symmetry breaking DQPT indicate that expectation values of site-averaged and time-averaged 2-local observables may be robust to MPS truncation errors, even up to times where the state deviates considerably from the true state. For the DQPT spatial and temporal averaging played an important role in the rapid convergence with bond dimension truncation. This is not the case for the correlation length which depends on the correlation function between specific spins at a specific time. This paradigm thus allows us to explore further distinctions between microstate and macroproperties and the tractability of quantum simulation.

\begin{figure}[ht] %  fnn
   \centering
   %\subfigure[]
   {\includegraphics[width=0.48\textwidth]{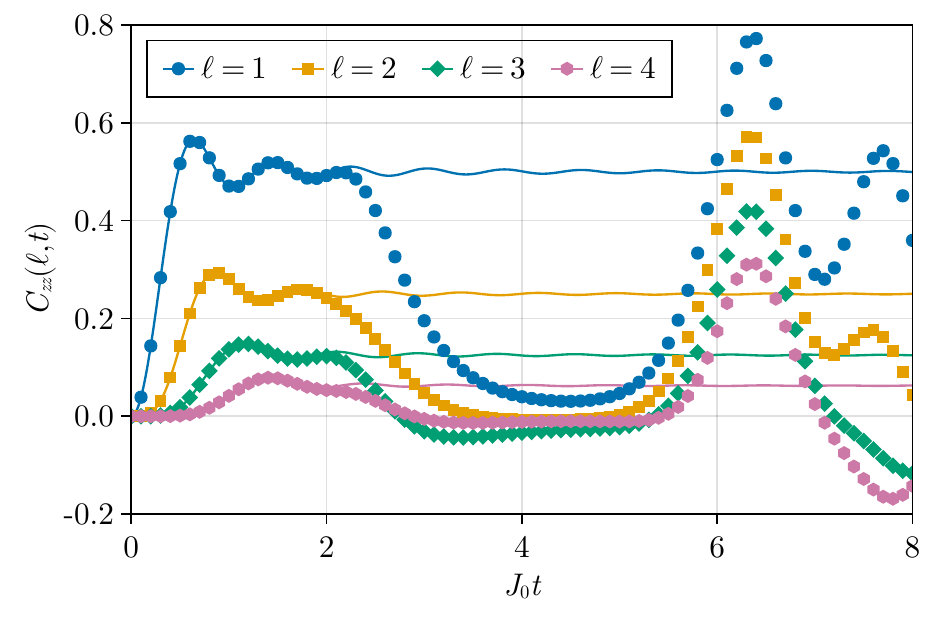}  } 
	\caption{Spatial correlation function in TFIM for $n=81$ spins when quenched to the critical point $B/J_0 = 1$, for different separations $\ell$ between two spins. The solid curves correspond to the exact results while the data points are for simulations with a bond dimension of 8. The step size used is $J_0 \Delta t = 10^{-2}$, but only data with a separation of 0.1 are shown. For early times ($J_0 t \leq 2$), the low bond dimension simulation tracks the exact results very well, but at later times it deviates dramatically from the exact values, calculated using the Gaussian Fermionic representation~\cite{derzhko1998numerical}.}
	\label{fig:CzzThermal}
\end{figure}

We consider here quenches from deep within the disordered phase ($B_{\mathrm{i}}/J_0 \to \infty$) to the critical point $B_{\mathrm{f}}/J_0 = 1$, where initially all spins are polarized along the $x$-direction, $\ket{\psi(0)} = \ket{\uparrow_x}^{\otimes n}$. Using 4th order time evolving block decimation (TEBD) to simulate the dynamics, Fig.~\ref{fig:CzzThermal} shows the behavior of $C_{zz}(t, \ell)$ for fixed distance $\ell$ as a function of the simulation time $J_0 t$. We provide details of the simulations in Appendix~\ref{app:TEBD}. We observe that the equal-time correlation for each value of the site separation $\ell$ reaches a quasi-steady-state value, with smaller separations reaching it sooner than larger $\ell$ values. As we will see, we only need a fixed range of $\ell$ to accurately extract the correlation length, and thus we will not need to scale the simulation time with $\ell$. 

Figure~\ref{fig:CzzThermal} additionally demonstrates how well a fixed bond dimension MPS simulation, in this case $\chi = 8$, does in approximating the exact correlation function. While the MPS simulation reproduces the correlation function at short times when the fixed bond dimension is sufficient to capture the exact state, the approximation fails at later times. Nonetheless, the long-time correlation length can be reliably extracted from the short-time quench dynamics that we can accurately simulate, well before the full many-body state has thermalized.

\begin{figure*}[ht] 
   \centering
   \subfigure[\ $J_0 t=2$]{\includegraphics[width=0.48\textwidth]{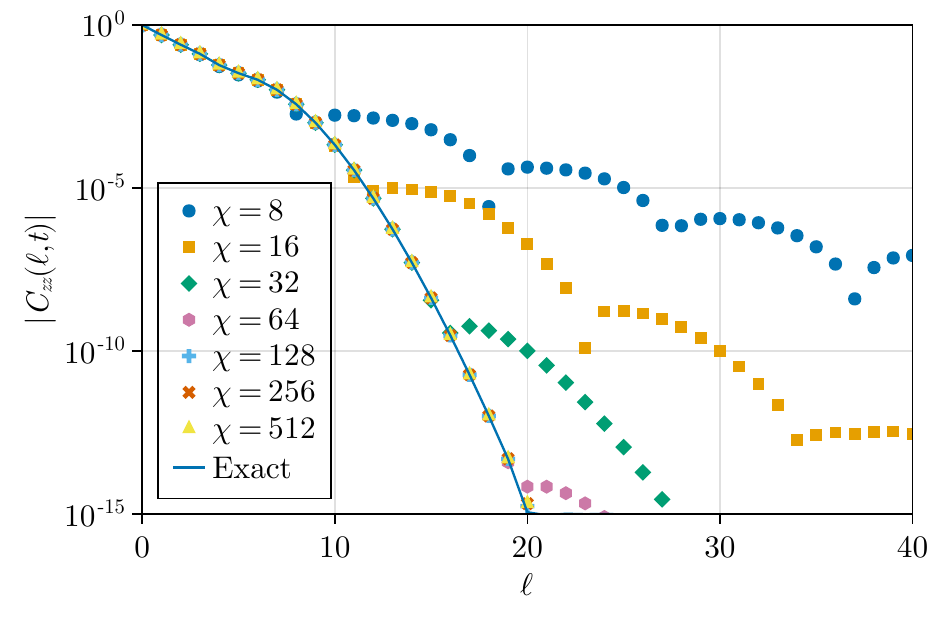}}
   \subfigure[\ $J_0 t=5$]{\includegraphics[width=0.48\textwidth]{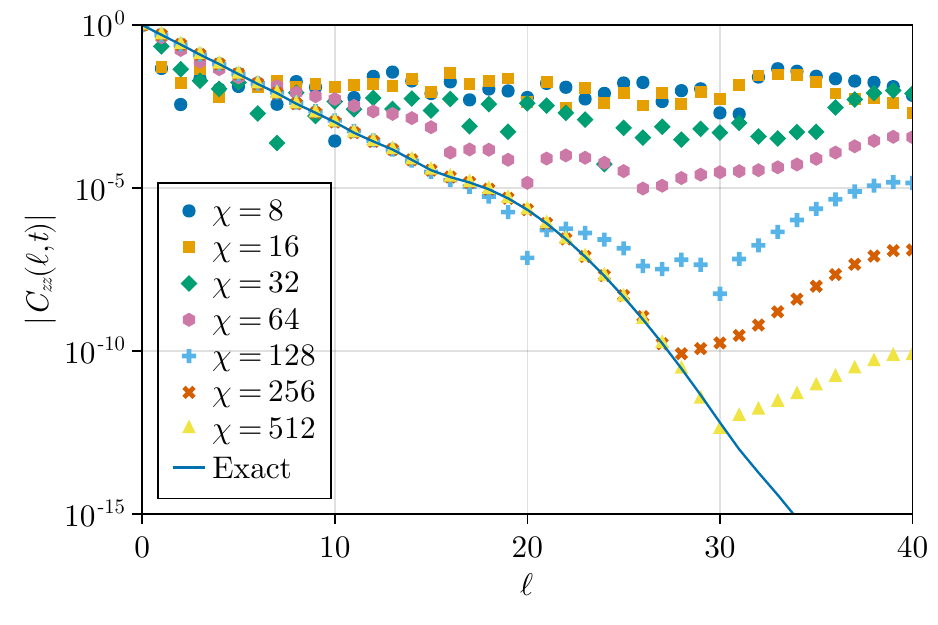}}
   \caption{Spatial correlation function in TFIM for $n=81$ spins when quenched to the critical point $B/J_0 = 1$, for different total quench times $J_0 t$. For small simulation times as in (a),  low bond dimension simulations can accurately reproduce the spatial correlation function for an extended $\ell$ range. For example, $\chi = 16$ is sufficient to accurately predict the behavior up to $\ell =10$. However for larger simulation times as in (b), only significantly higher bond dimension simulations are able to reproduce the spatial correlation function for the same range. For example, $\chi = 128$ is needed to accurately predict the behavior up to $\ell =10$. Exact simulations are done using the Gaussian Fermionic Hamiltonian representation~\cite{derzhko1998numerical}.} 
   \label{fig:Czz}
\end{figure*}

Figure~\ref{fig:Czz} illustrates the behavior of $C_{zz}(t, \ell)$ for two different fixed values of $J_0 t$ as we vary the separation distance $\ell$. One clearly observes a transition between two different exponential curves, with characteristic decay lengths $\xi_1$ at short separation $\ell$ and a weaker decay $\xi_2$ at large separation, as observed in~\cite{Kar2017}. The latter results from ``prethermalization" before correlations have spread to the entire system. By increasing the simulation time $J_0 t$, the range of separation distances~$\ell$ over which the first decay occurs can be made larger, and importantly $\xi_1$ is independent of time. In the infinite time limit $\xi_1$ defines pure exponential decay over the whole chain. Reproducing the full correlation function at large $\ell$, even at short times, requires a more accurate simulation using a larger bond dimension. However the short range behavior decay constant, which survives to
long times, is well approximated by a highly truncated~MPS.

\begin{figure}[h] %  f
   \centering
    \includegraphics[width=0.48\textwidth]{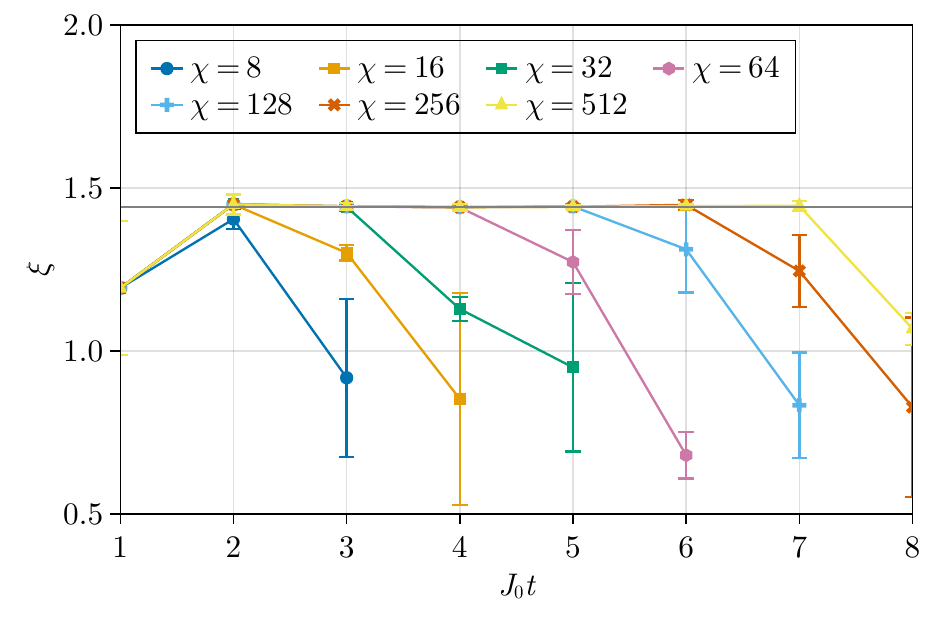}
    \caption{Results of fitting $|C_{zz}(t,\ell)|$ using the datapoints shown in Fig.~\ref{fig:Czz} at $\ell \leq 6$ to $\exp(-\ell/\xi)$. The thin black line corresponds to the long-time correlation length of $1/\ln(2)$. Error bars correspond to 95\% confidence interval. Moving from left to right after $J_0 t = 2$, the bond dimension must be doubled for every dimensionless unit of time increment in order to maintain an accurate estimate of the correlation length.}
   \label{fig:CorrelationLength}
\end{figure}

We show the correlation length $\xi_1$ extracted from the data in Fig.~\ref{fig:CorrelationLength}, which we obtain by fitting $|C_{zz}(t,\ell)|$ to $\exp(-\ell/\xi)$ using short-range correlations with $\ell \leq 6$ . If $J_0 t$ is too small $(J_0 t < 2)$, the extracted correlation length is not accurate even if the simulations are. However, already at $J_0 t=2$ we see that we can extract a very good estimate of the infinite-time correlation length, and we can do this using a relatively small bond dimension. Increasing $J_0 t$ does not practically improve our estimate of the infinite-time correlation length, but it does require simulations using a larger bond dimension to converge to the correct behavior. Specifically, increasing the dimensionless simulation time, $J_0 t$, by one unit requires us to increase the bond dimension by a factor of two in order to reproduce the correct correlation length. This can be seen in Fig.~\ref{fig:CorrelationLength}. At $J_0 t = 2$, bond dimensions  $\chi \geq 16$ reproduces the correct correlation length; at $J_0 t = 3$ only bond dimensions $\chi \geq 32$ reproduce the correct correlation length, etc. 

\begin{figure}[h] % 
   \centering
   \includegraphics[width=0.48\textwidth]{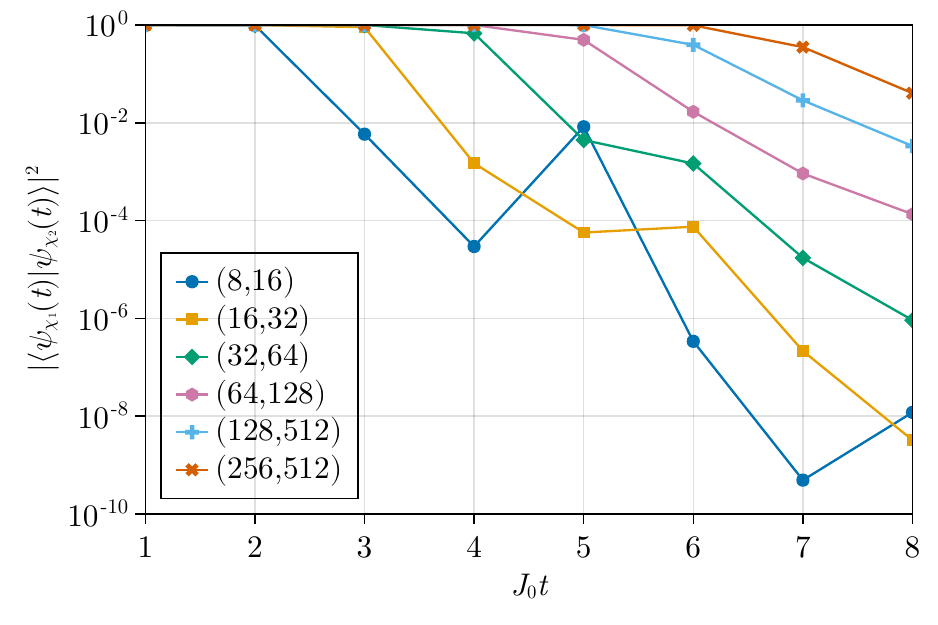} 
   \caption{Overlap squared of states simulated with different bond dimensions as a function of simulation time $T$ in the TFIM with $n=81$ spins quenched to the critical point $B/J_0 = 1$. The times at which each curve drops, which corresponds to when the two simulations no longer agree in terms of the many-body state with high fidelity, agrees with the times in Fig.~\ref{fig:CorrelationLength} when the estimate of the correlation length begins to fail.}
   \label{fig:CorrelationLengthOverlap}
\end{figure}

\begin{figure}[h] % 
   \centering
   \includegraphics[width=0.48\textwidth]{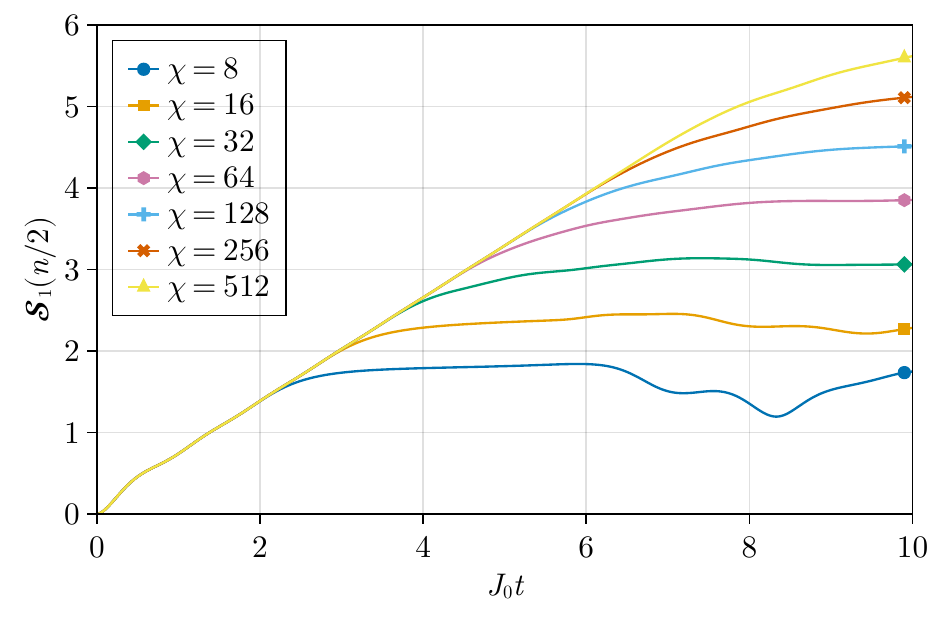} 
   \caption{Bipartite entanglement entropy between the two halves of the system for simulations	with different bond dimensions as a function of quench time $t$ in the TFIM for $n=81$ spins when quenched to the critical point $B/J_0 = 1$. The times at which each curve flattens, which corresponds to when the bond dimension is insufficient to capture the growth of entanglement, agrees with the times in Fig.~\ref{fig:CorrelationLength} when the estimate of the correlation length begins to fail.}
   \label{fig:CorrelationLengthEE}
\end{figure}

To further quantify the cost of simulation, in Fig.~\ref{fig:CorrelationLengthOverlap} we plot the squared overlap between states simulated with bond dimension $\chi$ and $2 \chi$. The time at which the simulated correlation length deviates from the true value for a fixed bond dimension $\chi$ agrees with the time at which the overlap between the full state with bond dimension $\chi$ deviates from that with bond dimension $2 \chi$. For example, the overlap of the states with bond dimensions $(16, 32)$ begins to deviate at $J_0 t \geq 3$, indicating that the simulation with bond dimension $\chi = 16$ is no longer accurate. This is simultaneously the value of $J_0 t$ for which the extracted correlation length deviates from the true value for $\chi = 16$. Similarly, the overlap of the states with bond dimensions $(32, 64)$ begin to deviate at $J_0 t \geq 4$, which coincides with the time for which the extracted correlation length deviates from the true value for $\chi = 32$. Before these times, the states from the simulations have overlap squared almost equal to 1. We note that these times also coincide with the times when the bipartite entanglement entropy between the two halves of the system saturates, as shown in Fig.~\ref{fig:CorrelationLengthEE}.

From these simulations we draw the following conclusions. In contrast to the DQPT, the correlation length does not involve spatial or time averaging. As such, it is more sensitive to the microstate.  For this model, one finds that accurate simulation of the correlation length requires accurate simulation of the full many-body state, which we see in the exponential scaling of the bond-dimension with time, $\chi \sim \mathcal{O}(2^{J_0 t})$. Formally, the correlation length is defined in the thermodynamic and infinite-time limits, and at the critical point the thermalized state exhibits volume-law entanglement, making classical simulation via MPS and TEBD intractable. For the 1D nearest-neighbor TFIM, one finds that the correlation length formally defined in the long-time limit is already available in the short-range short-time quench dynamics, making classical simulation tractable with a fixed bond dimension and run-times that scale as $\mathcal{O}(n \chi^3)$. The exact microstate in the thermodynamic limit is thus not necessary to extract the order parameter of interest.

%

% %%%%%%%%%%%%%%%%%%%%%%%%%%%%%%%%%%%%%%%%%%%%%%%%%%%%%%%%%%%%%%%%%%%%%%%%%%%%%%
\section{Discussion: The Role of Chaos}
\label{sec:RoleChaos}
%%%%%%%%%%%%%%%%%%%%%%%%%%%%%%%%%%%%%%%%%%%%%%%%%%%%%%%%%%%%%%%%%%%%%%%%%%%%%%
%
Random dynamics corresponding to evolution under chaotic Hamiltonians act as information scramblers, where local observables rapidly thermalize and information about the initial condition is spread throughout the system~\cite{Hayden2007, Rob2017}. As discussed in Sec.~\ref{sec:ModelsAndMethods}, the regime of interest for a quantum advantage in simulating local expectation values is where the full many-body state reaches volume-law entanglement, making classical simulation intractable, but where there is not complete scrambling to a typical quantum state~\cite{Leb1993, Gol2006, Leb2007, Bar2009, Dub2012, San2012, Fac2015, d2016quantum, Mi2022}, which would render the local state near maximally mixed. In order to better understand the findings of Sec.~\ref{sec:DynamicalQuantumPhaseTransition} and \ref{sec:CorrelationLengthFromQuench} in this context, we consider the approximation to local reduced density operators using  the framework of chaos and chaotic systems and how it affects the growth of entanglement and equilibration of local reduced density operators. In what follows, we focus on the case of $\alpha \to \infty$ for clarity. 

In addition to the mean adjacent level spacing ratio (Appendix~\ref{app:EnergyLevelSpacing}), the entanglement entropy of a bipartition of the Hamiltonian eigenvectors is another measure of chaoticity of the Hamiltonian~\cite{Lakshminarayan2001, karthik2007entanglement, Oma2022}. For chaotic Hamiltonians, most of the eigenvectors near the middle or bulk of the spectrum have volume-law scaling of entanglement entropy, and area-law scaling entanglement entropy for some states at the edges of the spectrum. For integrable Hamiltonians, there are eigenvectors with area-law scaling of entanglement entropy throughout the spectrum. We use this measure as it allows us to relate the chaoticity of the Hamiltonian to the entanglement entropies of the local reduced density operators in quantum quenches, which plays a role in determining the simulability of the local observables we have considered in Sec.~\ref{sec:DynamicalQuantumPhaseTransition} and ~\ref{sec:CorrelationLengthFromQuench}.

\begin{figure*}[h!] % 
    \centering
    \subfigure[]
    {\includegraphics[width=0.46\textwidth]{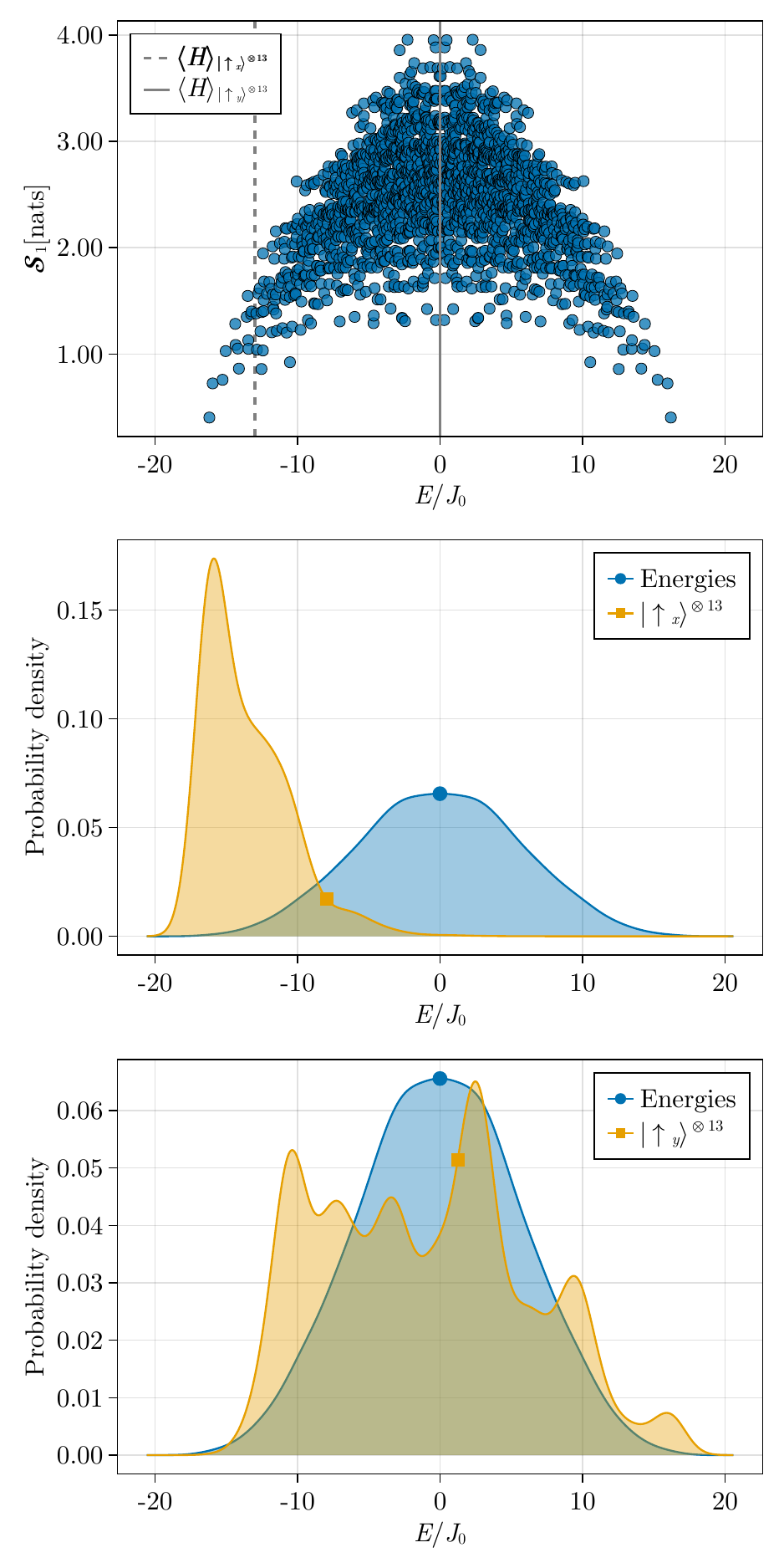} \label{fig:EigenvectorEntropiesPopulationTFIM}}
    \subfigure[]
    {\includegraphics[width=0.46\textwidth]{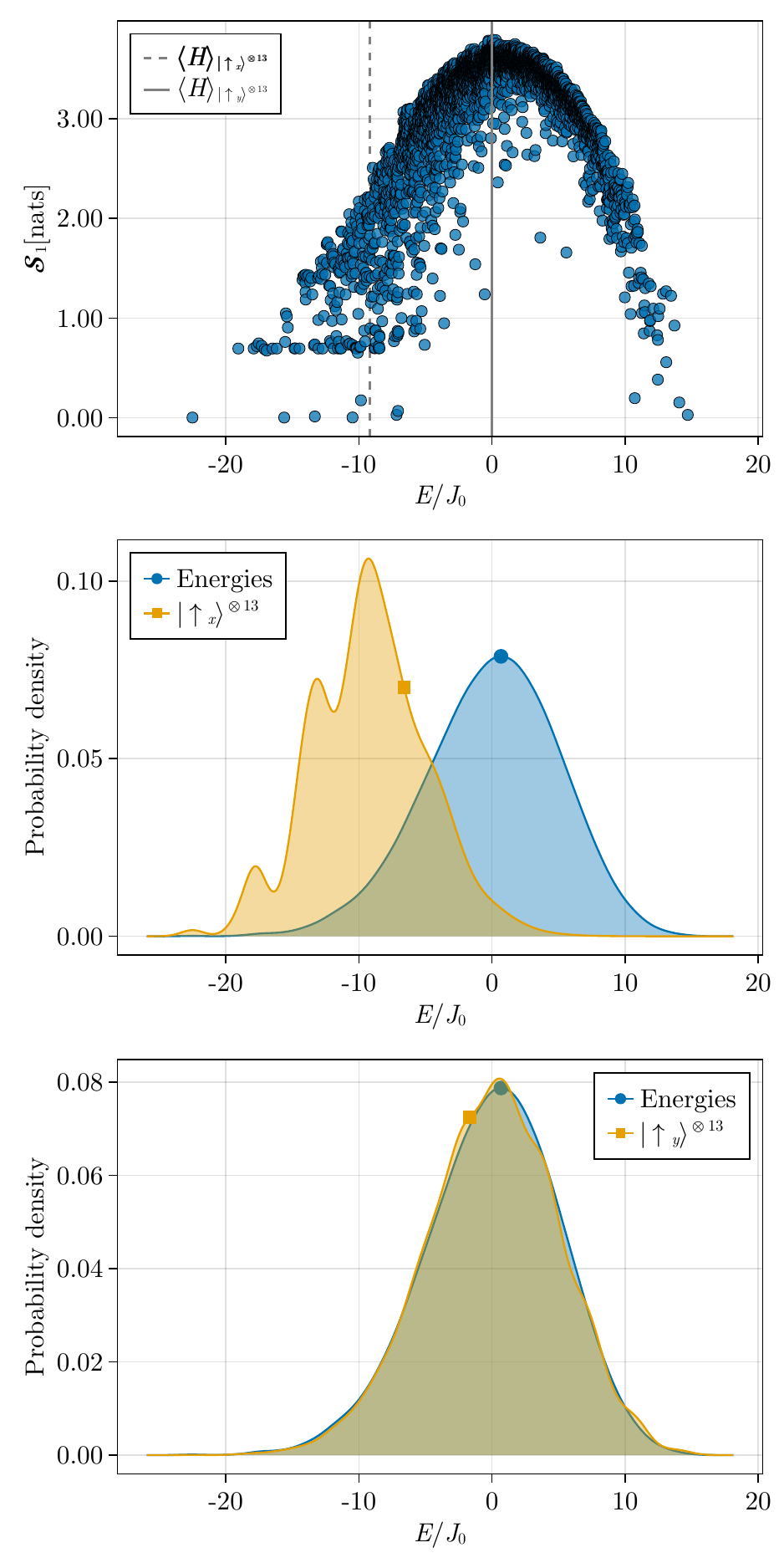}\label{fig:EigenvectorEntropiesPopulationSFIM}}
    \caption{Entanglement in TFIM and SFIM Hamiltonian eigenvectors and the population in energy eigenstates for different initial states. Top row: half-chain entanglement entropy, $\mathcal{S}_1$, in nats, of energy eigenvectors as a function of energy for (a) TFIM (b) SFIM with $\theta = \pi/4$. Vertical lines show the expectation value of the Hamiltonian in the states $|\uparrow_x\rangle^{\otimes 13}$ and $|\uparrow_y\rangle^{\otimes 13}$. Middle row: energy density of the initial state $| \uparrow_x \rangle^{\otimes 13}$ in the quenched eigenbasis and the Hamiltonian as a function of energy for (a) TFIM (b) SFIM with $\theta = \pi/4$. Bottom row: Energy density of the initial state $| \uparrow_y \rangle^{\otimes 13}$ in the quenched eigenbasis and the Hamiltonian as a function of energy for (a) TFIM (b) SFIM with $\theta = \pi/4$. The energy densities are calculated using kernel density estimation~\cite{sheather2004density, wkeglarczyk2018kernel} with Gaussian kernels whose widths are determined by Silverman's rule~\cite{silverman1981using, silverman1986density}.}
\end{figure*}

The TFIM studied in Sec.~\ref{sec:CorrelationLengthFromQuench} is an integrable model, so to better understand the role of chaos we break the spin-flip symmetry by slanting the magnetic field in the $z-x$ plane to an angle $\theta$ with the $z$-axis. The Hamiltonian now reads
\begin{equation} 
\begin{aligned}
  H = &
  - J_0 \sum_{\ell} \sigma^z_{\ell} \sigma^z_{\ell + 1}
  - B \sum_{\ell} \left( \cos(\theta)
  \sigma^z_{\ell} + \sin(\theta) \sigma^x_{\ell} \right).
  \label{eq:HamiltonianPowerLawSFIM}
\end{aligned}
\end{equation}
We call this the slanted field Ising model (SFIM), as studied in~\cite{prosen2007efficiency, leviatan2017quantum, rakovszky2022dissipation, vonkeyserlingk2022operator, cotler2023emergent, choi2023preparing, Mil2022}. When $\theta =0$ or $\theta =\pi$ the model reduces to the classical Ising model, and for $\theta = \pi/2$ it is the TFIM from Eq.~\eqref{eqt:alphaInftyTFIM}. To illustrate how different the properties of the spectra of these models are, we plot in the top row of Fig.~\ref{fig:EigenvectorEntropiesPopulationTFIM} and Fig.~\ref{fig:EigenvectorEntropiesPopulationSFIM} the bipartite half-chain von Neumann entanglement entropies of the eigenvectors of the Hamiltonian as a function of energy for the TFIM and SFIM with $B/J_0 = 1$. For the integrable TFIM, there are low entanglement eigenvectors throughout the spectrum, while for the chaotic SFIM most states near the bulk of the spectrum are highly entangled and states near the edges of the spectrum have low entanglement. Therefore, we generically expect larger entanglement entropy generation in the quench for the chaotic SFIM. By tuning $\theta$ we can probe the role of chaos and equilibration in the relative ease of approximating local expectation values. 

\begin{figure}[ht] %  f
    \centering
    \includegraphics[width=0.48\textwidth]{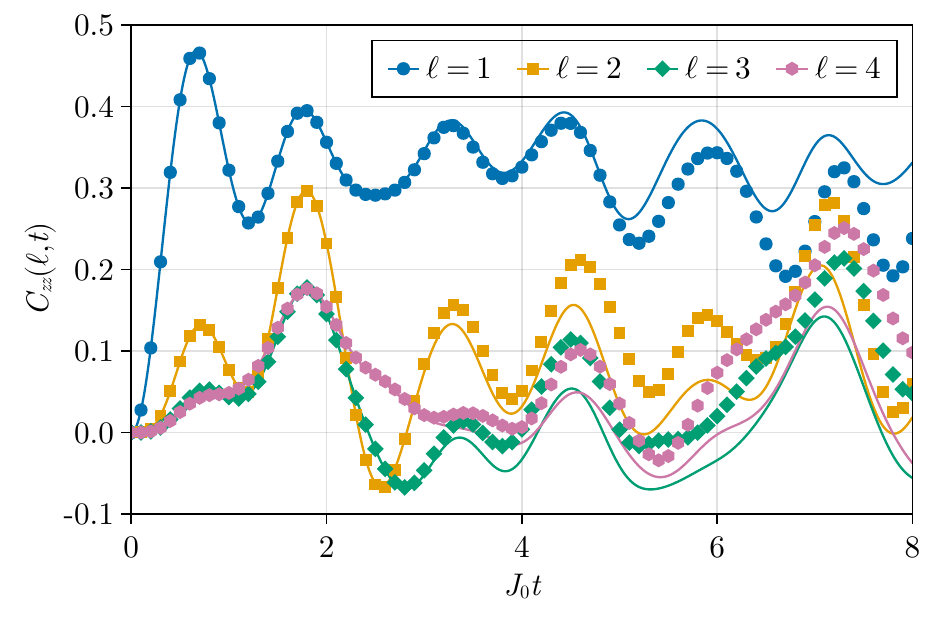} 
    \caption{Spatial correlation function in the SFIM with $\theta=\pi/4$, [Eq.~\eqref{eq:HamiltonianPowerLawSFIM}] with $n = 81$ spins when quenched to the critical point $B/J_0 = 1$, for different separations between two spins. The solid curves correspond to results using a maximum bond dimension of $512$ while the data points are for simulations with a bond dimension of 8. The step size used is $J_0 \Delta t = 10^{-2}$, but only data with a separation of 0.1 are shown. In contrast to Fig.~\ref{fig:CzzThermal}, the low bond dimension simulation approximates the correct behavior for longer times.}\label{fig:CzzHz}
\end{figure}

Consider the spin-spin correlation function for quenches from $|\uparrow_x\rangle^{\otimes n}$, as studied in Sec.~\ref{sec:CorrelationLengthFromQuench}, but now for the case of the SFIM with $\theta = \pi/4$. In Fig.~\ref{fig:CzzHz} we show the correlation functions simulated with bond dimensions $\chi = 8$ and $\chi = 512$, using TEBD as considered in Sec.~\ref{sec:CorrelationLengthFromQuench} with details in Appendix~\ref{app:TEBD}. We see that the estimates of the correlation functions agree for a significantly longer time compared to the integrable case of $\theta = \pi/2$ studied in Sec.~\ref{sec:CorrelationLengthFromQuench}. Thus, with the introduction of a nonzero longitudinal field, the system becomes chaotic, and we find that lower bond dimension simulations are better able to approximate the results of higher bond dimension simulations.

\begin{figure}[ht]
  \centering
  \includegraphics[width=0.48\textwidth]{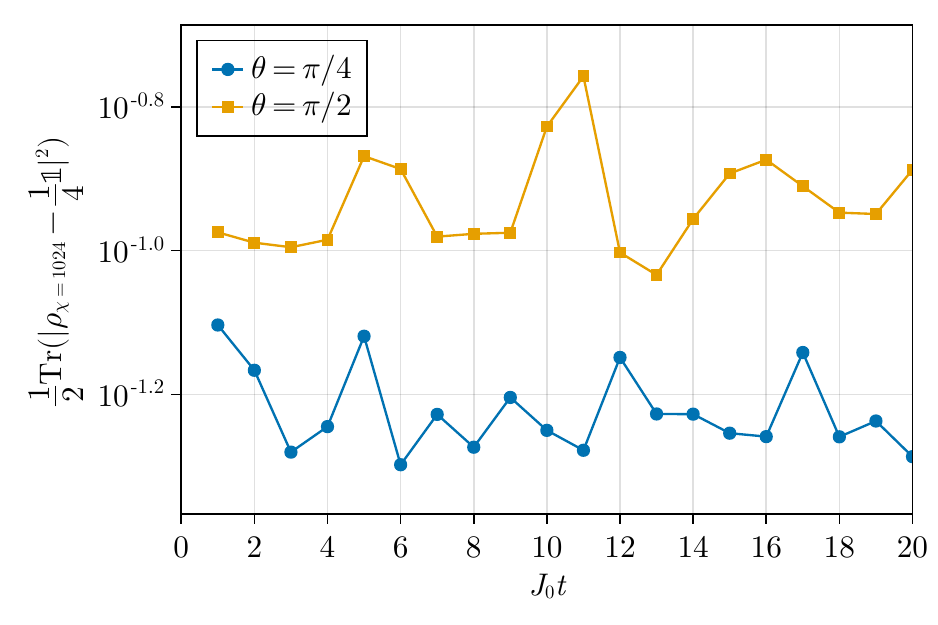}
  \caption{Comparison of the proximity of two-spin reduced density operator to the maximally mixed state attained for the reduced states in the TFIM (circles) and SFIM with $\theta = \pi/4$ (squares).  Here we choose $n=20$ spins and an initial state, $|\uparrow_x\rangle^{\otimes 20}$, for a Hamiltoninan quenched to $B/J_0=1$. Plotted is the squared HS distance between the two-spin reduced density operator for nearest neighbor spins at the center of the chain evaluated with bond-dimension $\chi=1024$ and the maximally mixed two-spin state as a function of time during quenches from $|\uparrow_x\rangle^{\otimes 20}$ to TFIM and $\theta = \pi/4$. The reduced density operator is closer to the maximally mixed case for the SFIM than the TFIM.}
  \label{fig:HSDistanceSFIMXPolarizedQuench}
\end{figure}

In order to understand the increased robustness of expectation values to truncation, we study the correlation between this and the closeness of the reduced density operator to the maximally mixed state. To illustrate this  we exactly simulate the TFIM and SFIM for $n = 20$, and in Fig.~\ref{fig:HSDistanceSFIMXPolarizedQuench} we plot the squared HS distance Eq.~\eqref{Eq:HS-Distance} between the two-body reduced density operator for nearest neighbors at the center of the chain and the maximally mixed state. For the initial state $|\uparrow_x\rangle^{\otimes n}$, the two-spin reduced density operator  is closer to the two-spin maximally mixed state for the $\theta = \pi/4$ SFIM than for the integrable model. These simulations therefore suggest an important relationship between the chaoticity of the quenched Hamiltonian and the efficiency with which we can simulate local expectation values via a truncated MPS representation. 

\begin{figure}[ht]
  \centering
  \includegraphics[width=0.48\textwidth]{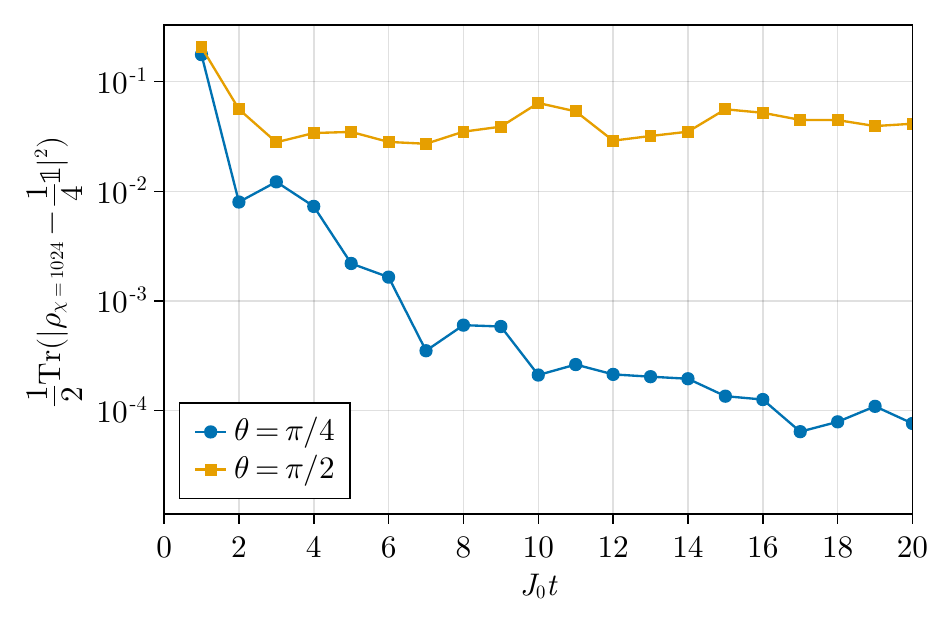}
  \caption{Same as Fig.~\ref{fig:HSDistanceSFIMXPolarizedQuench} but for an initial state $|\uparrow_y\rangle^{\otimes 20}$.  In contrast to that figure, the reduced density operator of the SFIM with this initial condition is significantly closer to the maximally mixed state in comparison to the TFIM  as this initial state thermalizes closer to a quantum typical state.}
  \label{fig:HSDistanceSFIMYPolarizedQuench}
\end{figure}

In addition to the Hamiltonian, the initial state plays an important role in the complexity of simulating quench dynamics. A chaotic Hamiltonian and an initial state with more support on energy eigenstates in the bulk of the spectrum, which are highly entangled, lead to higher effective temperature quenches with local reduced density operators being closer to the appropriate maximally mixed state. On the other hand, we expect nongeneric behavior for initial conditions with an average energy near the edges of the spectrum. For example, the $|\uparrow_x \rangle^{\otimes n}$ initial condition has an extensive average energy arising from the transverse field and is primarily supported in the low energy part of the spectrum for both the TFIM and $\theta = \pi/4$ SFIM (middle row of Figs.~\ref{fig:EigenvectorEntropiesPopulationTFIM} and ~\ref{fig:EigenvectorEntropiesPopulationSFIM}). We can therefore expect a low effective temperature for the long-time evolved state. In contrast, the $|\uparrow_y \rangle^{\otimes n}$ initial condition has an average energy of zero and is supported throughout the spectrum of the TFIM and $\theta = \pi/4$ SFIM (bottom row of Figs.~\ref{fig:EigenvectorEntropiesPopulationTFIM} and ~\ref{fig:EigenvectorEntropiesPopulationSFIM}). We can therefore expect a significantly higher effective temperature for this initial state. This is borne out when we consider a quench with this initial state (Fig.~\ref{fig:HSDistanceSFIMYPolarizedQuench}), where the difference between the integrable TFIM and chaotic SFIM is clearer -- the two-spin reduced density operator is significantly closer to the maximally mixed state for the chaotic SFIM than for the TFIM. 
%

%
%%%%%%%%%%%%%%%%%%%%%%%%%%%%%%%%%%%%%%%%%%%%%%%%%%%%%%%%%%%%%%%%%%%%%%%%%%%%
\section{Summary and Outlook}
\label{sec:SummaryOutlook}
%%%%%%%%%%%%%%%%%%%%%%%%%%%%%%%%%%%%%%%%%%%%%%%%%%%%%%%%%%%%%%%%%%%%%%%%%%%%
%
In this work we studied the extent to which we can classically simulate local expectation values in the context of quench dynamics of 1D Ising models using truncated MPS. We focused on whether a poor approximation to the exact many-body state may still yield an excellent approximation to the local expectation values, and characterize this as distinction between microstate and macroproperties. The microstate is specified by the full many-body quantum state, which can become intractable to simulate at long times and large enough system sizes when it becomes volume-law entangled. The macroproperties on the other hand correspond to the order parameters, which can be consistent with a multitude of microstates.

As our first example we considered the quench dynamics of the 1D TFIM with long range interactions in order to study the critical properties of DQPTs. The Hamiltonian of the TFIM has a $\mathbb{Z}_2$ symmetry that is explicitly broken by the initial state, and one defines two dynamical phases in which the symmetry is either restored or remains broken by the long-time dynamics. These phases are characterized by local order parameters involving spatial and temporal averaging of expectation values of one-spin and two-spin observables, and we find that they can be distinguished using low bond dimension MPS simulations. In particular we find that even when the approximation of the full many-spin state is poor, the approximation of the local reduced density operators can be orders of magnitude better. We used our simulations to also estimate the critical point and critical exponents of the phase transition using finite-size scaling analyses, and we found these critical properties to be insensitive to the MPS bond-dimension. We therefore conclude that the macroproperties associated with the DQPTs of 1D TFIM's can be well approximated with truncated MPS even if the microstates cannot.

For the simulations presented in our manuscript, we found that the system sizes were sufficiently large and evolution times short enough to minimize edge effects.
%This was particularly evident in simulations with larger system sizes.
As a result, we anticipate that similar results will also hold true for systems with periodic boundary conditions. While our MPS approach has provided valuable insights into dynamical critical phenomena in Ising models with open boundary conditions at relatively short times, further investigation of dynamical critical phenomena remains an open and intriguing area of research.

We next considered the quench dynamics of the 1D nearest neighbor TFIM in order to estimate the infinite-time correlation length after a quench to the critical point of the ground state quantum phase transition. In this case the relevant correlation function is more sensitive to the microstate as it is defined by an expectation value of specific spins at a given time as a function of their separation. While the correlation length is defined by the infinite time and thermodynamic limit when correlations have spread throughout the long chain, in practice the desired correlation length can be extracted at surprisingly short times, as found in Ref.~\cite{Kar2017}. The dynamics are such that local observables equilibrate rapidly to a steady state of a generalized Gibbs ensemble~\cite{calabrese2012quantum2}, allowing one to extract the correlation length very quickly, well before volume-law entanglement for the quantum state is reached. This allowed us to extract the correlation length accurately with short-time MPS simulations using very small bond dimension. Long-time simulations however required exponentially growing (with time) bond dimension in order to keep a high accuracy estimate of the correlation length. In both cases, a faithful simulation of the full many-body state was required to accurately estimate the correlation length. Therefore, in contrast to our DQPT example, we find that the local properties of this quench requires an excellent approximation of the microstate when using MPS, which cannot be done accurately for long times.

We can understand the preceding results by the role of chaos and local equilibration according to the ETH in quench dynamics. The 1D TFIM with nearest-neighbor interactions is integrable, so equilibrated local density matrices retain certain structure which can depend sensitively on the full quantum state. This is in contrast to the fully chaotic models such as the SFIM that includes both transverse and longitudinal $B$-fields. The chaotic SFIM Hamiltonian we study leads to more entanglement generation during the quench than the TFIM, and the equilibrated local state is the maximally mixed state, which has no useful information, and leads to better approximations of local expectation values with truncated MPS. 

Our results reveal a somewhat counterintuitive relationship between quantum chaos and the efficiency of classical simulation of expectation values of local observables in quench dynamics. The more chaotic the Hamiltonian, as measured, e.g., by level statistics, or entanglement entropy of its eigenvectors, the more resilient are these simulations to MPS bond-dimension truncation. Local expectation values in quench dynamics are easier to approximate for highly chaotic systems because at short times there is little entanglement and at longer times the system thermalizes such that the marginal states are close to maximally mixed. For such chaotic systems there is little information of interest in the marginal states. Our conclusion is opposite to what one might draw when considering other representations, such as in the Heisenberg picture in evolution of operators, where an operator evolves into an admixture of exponentially many operators with time, in different operator bases~\cite{parker2019universal, patramanis2022probing, heveling2022numerically}. Indeed, the operator spreading seen in operator entanglement of the SFIM led Prosen and {\ifmmode \check{Z}\else \v{Z}\fi{}nidari\ifmmode \check{c}\else \v{c}\fi{}} to ask, ``Is the efficiency of classical simulations of quantum dynamics related to integrability?'' with the reverse conclusion~\cite{prosen2007efficiency}. The choice of representation is thus essential when drawing conclusions about classical tractability. 

On the other hand, in the context of foundations of statistical mechanics and the justifications for ensembles thereof, the fact that ``typical" pure quantum states are compatible with predictions of statistical mechanics has been pointed out in the form of ETH or typicality~\cite{pop06ent, gol06can, rei07typ, sug13can, alba2015eigenstate, khodja2015relevance, d2016quantum, deutsch2018eigenstate,dymarsky2018subsystem, murthy2019bounds, noh2021eigenstate}. 
From this perspective, chaotic time evolution leading to more robust simulability may not appear as surprising, since such nonintegrability will more likely lead the state to become a random ``typical" state, which is arguably the very ingredient that enables a powerful effective theory such as statistical mechanics to emerge. 

As we suggested in Sec.~\ref{sec:Introduction}, useful quantum simulation of local order parameters in quench dynamics will require well-chosen problems whereby there is sufficient scrambling to render the problem intractable by classical simulation, but not enough scrambling to erase all information of interest in the local observables. We have focused our study of robustness to MPS truncation on problems involving estimating equal-time local correlation functions, and it would be important to understand whether one would observe similar robustness for two-time correlations, as in the study of out-of-time-ordered correlation functions (OTOCs). The success of MPS simulations in 1D, including modified approaches to truncation such as to extract long-time hydrodynamic properties \cite{vonkeyserlingk2022operator}, suggests that practical quantum advantages are more likely to be had in higher spatial dimensional systems \cite{daley2022practical}.

Another area where quantum simulation of quench dynamics may show a quantum advantage is in the study of local transport of globally conserved quantites in a many-body system through hydrodynamic and diffusion models. However, in a similar spirit to our work, transport properties for certain models may be accessible through appropriately chosen representations, such as truncated MPS~\cite{leviatan2017quantum} and the Heisenberg evolution of operators of interest~\cite{vonkeyserlingk2022operator, rakovszky2022dissipation}, even when description of the microscopic details of the dynamics is intractable. A key open question is the precise conditions under which these truncation schemes break down, and local correlation functions can not be efficiently simulated classically. 

In this work we restricted our attention to closed quantum systems. The presence of noise will naturally modify the probability distribution of measurements and affect the output of a NISQ device and the potential tractability of quantum simulation. There have been a number of related works which study how truncated descriptions of the microstate can reliably model the output to within the errors expected by noise. Zhou {\em et al.} showed that for realistic errors, a truncated MPS could efficiently simulate the microstate of 1D random circuits in a manner such that the truncation leads to an infidelity that is comparable with the gate infidelity of the noisy two-qubit gates \cite{zhou2020limits}. More detailed realistic modeling of decoherence using open quantum systems matrix product operators (MPOs) by Noh {\em et al.} \cite{noh2020efficient} and matrix product density operators (MPDOs) by Cheng {\em et al.} \cite{cheng2021simulating} showed similar results and set a bar for when random circuit sampling in 1D can achieve quantum advantage. The MPDO approach can accurately simulate the true probability distribution of a noisy quantum circuit when the noise is sufficiently large. More recently, Aharonov {\em et al.} have shown that there is a polynomial-time classical algorithm with which one can sample the output distribution of a noisy random quantum circuit for sufficiently deep circuits with a constant depolarizing error rate per gate \cite{aharonov2022polynomial, schuster2024}. Their method, similar to~\cite{vonkeyserlingk2022operator}, uses a truncation of ``Pauli paths,'' and a corresponding truncation of operator entanglement \cite{aharonov2022polynomial} when viewed in the Heisenberg picture. Approaches based on truncating Pauli paths has been successfully realized \cite{Fontana2023,Rudolph2023,Shao2023} to reproduce the results of a recent experiment of Floquet dynamics on a gate-based quantum device~\cite{kim2023evidence}. 

The general question remains -- what is the relationship between the robustness of a quantum simulation to noise and the complexity of the simulation. Quantum simulation is seen as a potential near-term goal for NISQ devices without the need for full fault-tolerant error correction because of the robustness of order parameters describing phases of matter. Our preliminary studies here show that such macroproperties can have an efficient description as a highly truncated MPS in closed quantum systems. Phenomena that require a more microscopic representation are likely to be more complex, but then they are likely to be more sensitive to noise and decoherence. A more complete study of the efficiency with which we can represent {\em open quantum systems} will be essential in order to demarcate the quantum-to-classical transition and its implications for achieving a useful quantum advantage in quantum simulation.

\begin{acknowledgments}
  We thank Pablo Poggi, Manuel Muñoz-Arias, Karthik Chinni, Sivaprasad Omanakuttan, Samuel Slezak, Frank Pollmann, Miles Stoudenmire, and Hoang Van Do for helpful discussions. We thank the UNM Center for Advanced Research Computing, supported in part by the National Science Foundation, for providing the research computing resources used in this work. This material is based upon work supported by the Air Force Office of Scientific Research under award number FA9550-22-1-0498 and FA9550-20-1-0123, work supported by the National Science Foundation Grant No. PHY-2116246, and is based upon work partially supported by the U.S. Department of Energy, Office of Science, National Quantum Information Science Research Centers, Quantum Systems Accelerator.
  We acknowledge the original people of New Mexico -- Pueblo, Navajo, and Apache, on whose traditional homelands the University of New Mexico stands. We acknowledge that the University of Oklahoma stands on traditional home of the Caddo Nation and Wichita and Affiliated Tribes and the hunting ground, trade exchange point, and migration route for the Apache, Comanche, Kiowa and Osage Nations.
\end{acknowledgments}

\appendix
%
%%%%%%%%%%%%%%%%%%%%%%%%%%%%%%%%%%%%%%%%%%%%%%%%%%%%%%%%%%%%%%%%%%%%%%%%%%%%%%
\section{Energy level spacing statistics for TFIM} \label{app:EnergyLevelSpacing}
%%%%%%%%%%%%%%%%%%%%%%%%%%%%%%%%%%%%%%%%%%%%%%%%%%%%%%%%%%%%%%%%%%%%%%%%%%%%%%
%

In this appendix, we assess how chaotic the TFIM Hamiltonians described by Eq.~\eqref{eq:HamiltonianPowerLawTFIM} are. Quantum chaos is one measure of complexity (not equivalent to integrability) that can affect the tractability of classical simulation of a given model. To assess the chaoticity of a model, we can consider the mean adjacent level spacing ratio, $\bar{r}$, which quantifies level repulsion of the energy eigenvalues, defined as \cite{atas2013distribution}
\begin{equation}
\begin{aligned}
  \bar{r} =& \textsc{Mean}(r_{k}),\label{eq:MeanAdjacentLevelSpacingRatio}
  \\
  r_{k} =&
  \frac
      {\min\left(\left({E_{k} - E_{k-1}}\right),
        \left(E_{k+1} - E_{k}\right)\right)}
      {\max\left(\left({E_{k} - E_{k-1}}\right),
        \left(E_{k+1} - E_{k}\right)\right)},
\end{aligned}
\end{equation}
where $E_k$ is the $k$-th lowest energy eigenvalue. An integrable Hamiltonian would correspond to Poissonian level spacing statistics with $\bar{r} \approx 0.386$, while a chaotic Hamiltonian, in this case, would correspond to $\bar{r}$ of the Gaussian orthogonal ensemble (GOE) of random matrix theory, with $\bar{r} \approx 0.535$ \cite{atas2013distribution}. Another measure of the chaoticity of a model is the entanglement entropy of the Hamiltonian eigenvectors, which we use in Sec.~\ref{sec:RoleChaos}.

\begin{figure}
    \centering
    \includegraphics[width=0.48\textwidth]{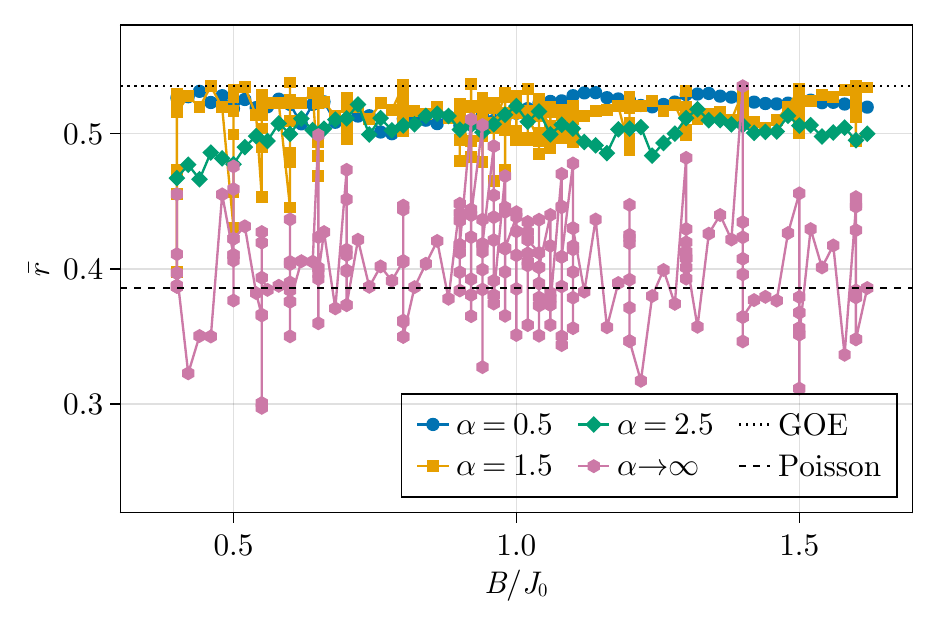}
    \caption{Mean adjacent level spacing ratio, $\bar{r}$, defined in Eq.~\eqref{eq:MeanAdjacentLevelSpacingRatio}, for the Hamiltonian in Eq.~\eqref{eq:HamiltonianPowerLawTFIM}, for a few values of $\alpha$, as functions of the magnetic field, $B$, for $n = 13$. The dotted line corresponds to $\bar{r} \approx 0.535$ for a typical chaotic Hamiltonian sampled from the Gaussian orthogonal ensemble and the dashed line corresponds to $\bar{r}\approx 0.386$ for a Hamiltonian with Poisson level spacing statistics.}
    \label{fig:MeanAdjacentLevelSpacingRatio}
\end{figure}

In Fig.~\ref{fig:MeanAdjacentLevelSpacingRatio} we plot $\bar{r}$ for TFIMs governed by Eq.~\eqref{eq:HamiltonianPowerLawTFIM} for several values $\alpha$ and $n = 13$ spins. We see high chaoticity, approaching the expected value for the GOE, for $0.5\le\alpha\le 2$. For large $\alpha$ we approach level-repulsion associated with the Poisson statistics of regular systems. This has implications for equilibration associated with quench dynamics, which is typically understood using versions of the eigenstate thermalization hypothesis (ETH)~\cite{alba2015eigenstate, khodja2015relevance, d2016quantum, deutsch2018eigenstate,dymarsky2018subsystem, murthy2019bounds, noh2021eigenstate}. 

%
%%%%%%%%%%%%%%%%%%%%%%%%%%%%%%%%%%%%%%%%%%%%%%%%%%%%%%%%%%%%%%%%%%%%%%%%%%%%%%%%
\section{Relating errors to Hilbert-Schmidt distance}
\label{app:Bounds}
%%%%%%%%%%%%%%%%%%%%%%%%%%%%%%%%%%%%%%%%%%%%%%%%%%%%%%%%%%%%%%%%%%%%%%%%%%%%%%%%
%

We show how to relate the difference in expectation values of observables estimated from two density matrices to the Hilbert-Schmidt (HS) distance between them. Let us consider an observable $\mathcal{A}$ whose expectation value is estimated from two density matrices $\rho_1$ and $\rho_2$ as
\begin{equation}
    \langle{\mathcal{A}}\rangle_{\rho_1} = \Tr\left(\rho_1 \mathcal{A}\right)
    ; \qquad
    \langle{\mathcal{A}}\rangle_{\rho_2} = \Tr\left(\rho_2 \mathcal{A}\right).
\end{equation}
For example, $\rho_1$ could be the reduced density operator calculated from a MPS with bond-dimension $\chi_1$, and $\rho_2$ could be the reduced density operator calculated from a MPS with bond-dimension $\chi_2$. The squared difference in the two estimates for the expectation value is given by
\begin{eqnarray}
    \big(\langle{\mathcal{A}}\rangle_{\rho_1} - \langle{\mathcal{A}}\rangle_{\rho_2}\big)^2
    &=& \big(\Tr\left(\rho_1 \mathcal{A}\right) - \Tr\left(\rho_2 \mathcal{A}\right)\big)^2 \nonumber \\
    &=& \big(\Tr\left( \left(\rho_1 - \rho_2\right) \mathcal{A} \big)\right)^2.
\end{eqnarray}
The last expression can be interpreted as the HS inner product or overlap between vectors $u \equiv \rho_1 - \rho_2$ and $v \equiv \mathcal{A}$
\begin{equation}
    \langle u, v \rangle
    = \Tr\left(u^\dagger v\right).
\end{equation}
Using the Cauchy-Schwarz inequality, 
\begin{equation}
    \left|\langle u, v \rangle\right|^2 \leq 
    \langle u, u \rangle
    \langle v, v \rangle,
\end{equation}
we have that
\begin{equation}
    \left|\Tr\left( u^\dagger v \right)\right|^2 \leq
    \Tr\left(u^\dagger u \right) \Tr\left( v^\dagger v \right).
\end{equation}

Since our operators are Hermitian, $u^\dagger = u$ and $v^\dagger = v$, we have
\begin{equation}
    \left|\Tr\left(u v \right) \right|^2 \leq
    \Tr\left(u^2 \right) \Tr\left(v^2 \right).
\end{equation}
Thus we have the inequality
e
\begin{eqnarray} \label{eq:ErrorBoundAppendix}
    \big(\Tr\left(\rho_1 - \rho_2 \right) \mathcal{A} \big)^2
    &\leq&
    \Tr\left(\mathcal{A}^2 \right) \Tr\left(\left(\rho_1 - \rho_2 \right)^2 \right)\nonumber \\
    &=& \Tr\left(\mathcal{A}^2 \right)
    \mathcal{D}_{\mathrm{HS}}^2 \left(\rho_1, \rho_2 \right),
\end{eqnarray}
where $\mathcal{D}_{\mathrm{HS}}^2 \left(\cdot, \cdot \right)$ is the squared HS distance between the operators:
\begin{equation} \label{eq:HSDistanceDefinitionAppendix}
    \mathcal{D}^2_{\mathrm{HS}}\left(\rho_1, \rho_2 \right) = \Tr\left(\left(\rho_1-\rho_2 \right)^\dagger \left(\rho_1-\rho_2 \right) \right).
\end{equation}
Thus we find that the squared difference in the two estimates of the observable is upper-bounded by the squared HS distance between their associated density operators.

Next, we derive a similar relationship for the  time- and site- averaged expectation value of an observable. We assume that $\mathcal{A}^{(j,k)}$ is a two-spin observable acting non-trivially only on sites $j$ and $k$ and is identical on all pairs of sites, i.e. its operator representation is identical $\forall j, k$, which we denote by $\mathcal{A}^{(2)}$. The time- and site- averaged expectation value of the observable is given by
\begin{equation}
  \overline{
    \left\langle \frac{1}{n^2} \sum_{j,k} \mathcal{A}^{(j,k)} \right\rangle}
  = \frac{1}{T} \int_0^T dt
  \frac{1}{n^2} \sum_{j,k} \Tr \left( \mathcal{A}^{(j,k)} \rho(t) \right),
\end{equation}
where
\begin{equation}
  \overline{\bullet} = \frac{1}{T} \int_0^T dt \left( \bullet \right),
\end{equation}
denotes the time-average over a time interval from $0$ to $T$.
 
We first consider the average 2-site error of the reduced density operator. Letting $\rho^{(j,k)}$ denote the reduced density operator on sites $j$ and $k$, we find that the error of the 2-site averaged error is given by
\begin{align}
&\left| { \left \langle \frac{1}{n^2}\sum_{j,k} \mathcal{A}^{(j,k)}(t) \right \rangle_{\chi_1}}  -  { \left \langle \frac{1}{n^2}\sum_{j,k} \mathcal{A}^{(j,k)}(t) \right \rangle_{\chi_2}}  \right|^2 \nonumber\\
&\;\;= \left|  \mathrm{Tr} \left(  \mathcal{A}^{(2)}  \frac{1}{n^2}\sum_{j,k} \left(\rho^{(j,k)}_{\chi_1}(t) - \rho^{(j,k)}_{\chi_2}(t) \right) \right) \right|^2 \nonumber\\
&\;\;\leq  \mathrm{Tr} \left( \left|  \mathcal{A}^{(2)} \right|^2 \right) \mathrm{Tr} \left( \left| \frac{1}{n^2}\sum_{j,k} \left( \rho^{(j,k)}_{\chi_1}(t) - \rho^{(j,k)}_{\chi_2}(t) \right) \right|^2\right) \nonumber\\
&\;\;=4 \mathcal{D}^2_{\mathrm{HS}} \left( {\frac{1}{n^2}\sum_{j,k} \rho^{(j,k)}_{\chi_1} (t)},\; \frac{1}{n^2}{\sum_{j,k}\rho^{(j,k)}_{\chi_2} (t)} \right). \label{eqt:Case2}
\end{align}
We can now easily consider the role of time-averaging by repeating the same calculation with the time-average of the expectation values:
\begin{align}
&\frac{1}{n^2} \sum_{j,k} \left| \overline{ \left\langle \mathcal{A}^{(j,k)} \right\rangle}_{\chi_1}  -  \overline{ \left\langle  \mathcal{A}^{(j,k)} \right\rangle}_{\chi_2}  \right|^2 \nonumber\\
&\;\;\;\;\;\;\leq \frac{4}{n^2}   \sum_{j,k}  \mathcal{D}^2_{\mathrm{HS}} \left( \overline{\rho^{(j,k)}_{\chi_1}(t)}, \overline{\rho^{(j,k)}_{\chi_2}(t)} \right). \label{eqt:Case3}
\end{align} 
The above calculation corresponds to first calculating the time-averaged 2-site reduced density operator, then calculating the squared HS distance, and finally averaging over all sites. If we change the order to first averaging over sites and then over time before calculating the distance, we find that
\begin{align}
&\left| \overline{ \left\langle \sum_{j,k} \mathcal{A}^{(j,k)} \right\rangle}_{\chi_1}  -  \overline{ \left\langle \sum_{j,k} \mathcal{A}^{(j,k)} \right\rangle}_{\chi_2}  \right|^2 \nonumber\\
&\;\;\;\;\;\;\leq 4 \mathcal{D}^2_{\mathrm{HS}} \left( \overline{\frac{1}{n^2}\sum_{j,k} \rho^{(j,k)}_{\chi_2}}, \overline{\frac{1}{n^2}\sum_{j,k}\rho^{(j,k)}_{\chi_1}} \right) \label{eqt:Case4}
\end{align}
The last term is the squared HS distance of the site- and time-averaged reduced density operator. Our expression provides an upper-bound for the error of the expectation value of space- and time-averaged observables in terms of the squared HS distance between the site- and time-averaged reduced density matrices.

The different inequalities derived in Eqs.~\eqref{eqt:Case2}-\eqref{eqt:Case4} correspond to whether we consider the site- and time-averaged reduced density operator. We use these different cases for the HS distance in Fig.~\ref{fig:HSD} of the main text.

%
%%%%%%%%%%%%%%%%%%%%%%%%%%%%%%%%%%%%%%%%%%%%%%%%%%%%%%%%%%%%%%%%%%%%%%%%%%%%%%%%
\section{Mean-field description of the DQPT}
\label{app:DQPTSimplePicture}
%%%%%%%%%%%%%%%%%%%%%%%%%%%%%%%%%%%%%%%%%%%%%%%%%%%%%%%%%%%%%%%%%%%%%%%%%%%%%%%%
%
A simple picture for DQPTs can be obtained by considering the $\alpha = 0$ case, which is a mean-field theory in the thermodynamic limit~\cite{zhang2017observationmany, de2023non, Zun2018, lang2018dynamical, chinni2021effect, chinni2022trotter, munoz2020simulation, munoz2021nonlinear, munoz2022floquet, munoz2023phase}. This is also a more computationally tractable case. The Hamiltonian in Eq.~\eqref{eq:HamiltonianPowerLawTFIM} can be expressed in terms of the total angular momentum operators, which simplify to powers of the average magnetization in the mean-field limit,
\begin{equation}
H = - B n 2 S_x - n J_0 \left( \left(2 S_z \right)^2 - \frac{1}{n} \ident \right),
\end{equation}
and it reduces to the LMG model \cite{Lip1965}. This Hamiltonian is invariant under permutation of the spins, and since the initial state $\ket{\uparrow_z}^{\otimes n}$ is permutation symmetric under spin permutation, the evolution under the Hamiltonian is restricted to the permutation symmetric subspace of dimension $n+1$. The polynomial scaling of the relevant Hilbert space means that we can perform exact simulations for significantly larger sizes $n$ and larger simulation times $J_0 t$.

Another advantage of the case $\alpha = 0$ is that it lends itself to a semi-classical treatment. In the semi-classical limit we can think of the Hamiltonian as giving rise to a potential energy \emph{density} profile described by
\begin{equation}
    V(\theta, \phi) = -B \sin(\theta) \cos(\phi) - J_0 \cos^2(\theta) ,
\end{equation}
which gives rise to a double-well potential. The parameter $B/J_0$ controls the height of the barrier between the two minima of the double well. In the limit $B/J_0 \to \infty$, the barrier vanishes. For $B / J_0 = 0$, we can associate the two minima with the two ferromagnetic ground states, $\ket{\uparrow_z \dots \uparrow_z}$ and $\ket{\downarrow_z \dots \downarrow_z}$. In the DQPT protocol, the initial state is one of these states (i.e. in one of the ground states of the Hamiltonian with $B/J_0 =0$ with a high barrier), and we quench the Hamiltonian to $B/J_0 \neq 0$ with a lower barrier.

The initial state determines the energy of the state, which remains conserved during the evolution. Suppose we start with the $\ket{\uparrow_z \dots \uparrow_z}$ state, with a positive magnetization. If the energy of this state is below the height of the barrier of the quenched Hamiltonian, the semiclassical evolution is the system bouncing back-and-forth within the \emph{same} well it starts in. Thus the magnetization stays positive as a function of time, and thus $M_{z} > 0$.

If the energy of the state is above the height of the barrier, then the semiclassical evolution will be the system bouncing back-and-forth between \emph{both} wells. Thus we expect the magnetization to oscillate between positive and negative values, and hence a vanishing time-averaged magnetization $M_{z} = 0$. (At finite $n$ the oscillation has an amplitude that decays to zero.) Thus there is a critical value of $B/J_0$ that separates the two phases given by when the energy of the initial state matches exactly the height of the barrier. This critical value turns out to be at $B/J_0 =1$ for $\alpha = 0$.

Why is the minimum of $M_{zz}$ a good indicator of the location of the phase transition? Suppose we approach the critical point from $B/J_0 <1$. While we expect $M_{z} > 0$ and the evolution to be periodic, the system spends a larger and larger fraction of the time in a single period near $\theta = \pi/2$ as $B/J_0 \to 1^{-}$. Thus, as we approach $B/J_0 \to 1^{-}$, not only does $M_{z}$ get smaller but so does $M_{zz}$.

Similarly, if we approach the critical point from $B/J_0 > 1$, we have $M_{z} = 0$ because the evolution spends equal amounts of time with $\cos \theta >0$ as $\cos \theta < 0$. For $B/J_0 \gg 1$, the wells are almost indistinguishable, and the system spends almost equal time in all regions and gives $M_{zz} \approx 1/2$. However, as we approach $B/J_0 \to 1^{+}$, the fraction of time the evolution spends near $\cos \theta = 0$ grows and hence $M_{zz}$ gets smaller. Thus we find a global minimum in $M_{zz}$ at the critical point.

We can be more formal about this description. The semiclassical equations of motion are given by the equations of motion of a classical unit-norm vector in 3-dimensional space (see for example Ref.~\cite{Owerre2015}):
\begin{subequations}
\begin{align}
\frac{1}{2} \sin \theta(t) \frac{d}{dt} \theta(t) &= {B} \sin \theta(t) \sin \phi(t) \\
- \frac{1}{2} \sin \theta(t) \frac{d}{dt} \phi(t) &= \cos\theta(t) \left( 2 J_0 \sin \theta(t) - B \cos \phi(t) \right).
\end{align}
\end{subequations}
It then follows that $d V / dt = 0$. For our initial condition with $\theta(0) = 0$, we have $V/J_0 = -1$, so the energy conservation condition gives us $\sin \theta(t) = B/J_0 \cos \phi(t)$. Using this condition, we can simplify our expression for the dynamics of the angle $\theta(t)$:
\begin{equation}
\frac{d}{dt} \theta(t) = \left\{ \begin{array}{lr}
2 J \sqrt{ \left(\frac{B}{J_0} \right)^2 - \sin^2 \theta(t) } \ , & \sin \phi(t)  > 0 \\
-2 J \sqrt{ \left(\frac{B}{J_0} \right)^2 - \sin^2 \theta(t) } \ , & \sin \phi(t)  < 0
\end{array} \right.
\end{equation}
This gives us the turning points of the semiclassical evolution when the evolution is constrained to a single well: for $B/J_0 < 1$, the minimum and maximum angles are $0$ and $\sin^{-1} (B/J_0)$.

In order to calculate $M_{z}$ and $M_{zz}$, it suffices to consider the values these time-averaged quantities take during a single half-period of the evolution: for $B/J_0 < 1$, $\cos \theta(t)$ and $\cos^2 \theta(t)$ have the same functional form for the first half-period as the second half-period; for $B/J_0 > 1$, $\cos \theta(t)$ in the second half-period has the opposite sign to the first half-period, whereas $\cos^2 \theta(t)$ has the same functional form. Therefore, we can write:
\begin{widetext}
\begin{eqnarray}
M_{z} = \left\{
\begin{array}{lr}
\frac{1}{T_{1/2}} \int_0^{T_{1/2}} \cos \theta(t) dt = \frac{\int_0^{\theta_t} \frac{\cos \theta}{\sqrt{\left( \frac{B}{J_0} \right)^2 - \sin^2(\theta)}} d\theta}{\int_0^{\theta_t} \frac{1}{ \sqrt{\left( \frac{B}{J_0} \right)^2 - \sin^2(\theta)}} d\theta} \ , & B/J_0 < 1\\
0 \ , & B/J_0 > 1
\end{array}\right.
\end{eqnarray}
\begin{eqnarray}
M_{zz} = \frac{1}{T_{1/2}} \int_0^{T_{1/2}} \cos^2 \theta(t) dt  = \left\{
\begin{array}{lr}
 \frac{\int_0^{\theta_t} \frac{\cos^2 \theta}{\sqrt{\left( \frac{B}{J_0} \right)^2 - \sin^2(\theta)}} d\theta}{\int_0^{\theta_t} \frac{1}{ \sqrt{\left( \frac{B}{J_0} \right)^2 - \sin^2(\theta)}} d\theta} \ , & B/J_0 < 1\\
 \frac{\int_0^{\pi/2} \frac{\cos^2 \theta}{ \sqrt{\left( \frac{B}{J_0} \right)^2 - \sin^2(\theta)}} d\theta}{\int_0^{\pi/2} \frac{1}{ \sqrt{\left( \frac{B}{J_0} \right)^2 - \sin^2(\theta)}} d\theta} \ , & B/J_0 > 1
\end{array}\right.
\end{eqnarray}
\end{widetext}
where $\theta_t = \sin^{-1}\left( \frac{B}{J_0} \right)$. We can now readily check that $M_{zz}$ attains a minimum as we take $B/J_0 \to 1$.

%
%%%%%%%%%%%%%%%%%%%%%%%%%%%%%%%%%%%%%%%%%%%%%%%%%%%%%%%%%%%%%%%%%%%%%%%%%%%%%%%%
\section{Details of MPS methods}
%%%%%%%%%%%%%%%%%%%%%%%%%%%%%%%%%%%%%%%%%%%%%%%%%%%%%%%%%%%%%%%%%%%%%%%%%%%%%%%%
%

In this appendix, we provide details of the MPS methods we used throughout the manuscript. We used the time-dependent variational principle for MPS~\cite{haegeman2011time, haegeman2016unifying} to simulate the quench in the long-range TFIM with $\alpha = 1.5$ as considered in Sec.~\ref{sec:DynamicalQuantumPhaseTransition}. For the 1D nearest-neighbor TFIM and SFIM considered in Sec.~\ref{sec:CorrelationLengthFromQuench} and Sec.~\ref{sec:RoleChaos}, we used the time-evolving block decimation method~\cite{vidal2003efficient, vidal2004efficient, daley2004time}.

%
%%%%%%%%%%%%%%%%%%%%%%%%%%%%%%%%%%%%%%%%%%%%%%%%%%%%%%%%%%%%%%%%%%%%%%%%%%%%%%%%
\subsection{Random matrix product states}
\label{app:RandomMPS}
%%%%%%%%%%%%%%%%%%%%%%%%%%%%%%%%%%%%%%%%%%%%%%%%%%%%%%%%%%%%%%%%%%%%%%%%%%%%%%%%
%
In Sec.~\ref{sec:DynamicalQuantumPhaseTransition} and Fig.~\ref{fig:HSD}, we consider random MPS, sampled from the Gaussian tensor ensemble. These MPS are sampled using the function \texttt{ITensors.randomMPS} of the ITensor library version 0.3~\cite{ITensor, ITensor-r0.3}. 
%Each tensor of the MPS in Eq.~\eqref{eq:MatrixProductState} is sampled from a Gaussian ensemble, with the number.
The dimensions of the tensors are up to $2 \times \chi \times \chi$ for MPS with bond-dimension $\chi$. We sampled MPS with bond dimension $\chi = 128$ and $\chi = 64$ and compared the two-site reduced density operators obtained from them, after site averaging and site and time averaging.

%
%%%%%%%%%%%%%%%%%%%%%%%%%%%%%%%%%%%%%%%%%%%%%%%%%%%%%%%%%%%%%%%%%%%%%%%%%%%%%%%%
\subsection{Time evolution algorithms}
\label{app:MPSTimeEvolution}
%%%%%%%%%%%%%%%%%%%%%%%%%%%%%%%%%%%%%%%%%%%%%%%%%%%%%%%%%%%%%%%%%%%%%%%%%%%%%%%%
%

Our work is focused on dynamics, therefore our primary aim is to represent the evolution of a quantum state as a series of transformations from one MPS representation into another, based on the time-independent Schrödinger equation (TDSE),
\begin{equation}
    \frac{\partial}{\partial t} | \psi \rangle = -\mathrm{i} H |\psi\rangle
    ,
    \label{eq:TimeDependentSchrodingerEquationAppendix}
\end{equation}
where we have set $\hbar \equiv 1$. The solution to this can be formally written as
\begin{equation}
    |\psi(t) \rangle
    = \exp\left(- \mathrm{i} t H \right)
    |\psi(t) \rangle
    ,
    \label{eq:TimeEvolutionOperatorAppendix}
\end{equation}
where $\exp\left(- \mathrm{i} t H \right)$ is the time evolution operator generated by the Hamiltonian $H$. We use two methods to approximate Eq.~\eqref{eq:TimeEvolutionOperatorAppendix} as applied to the MPS representation. For 1D nearest neighbor models considered in Sec.~\ref{sec:CorrelationLengthFromQuench} and Sec.~\ref{sec:RoleChaos}, we do this using the time-evolving block decimation (TEBD) method \cite{vidal2003efficient,vidal2004efficient, daley2004time, white2004real}, which involves a Trotterization of the time-evolution unitary as well as a truncation of the MPS. For long-range interacting models considered in Sec.~\ref{sec:DynamicalQuantumPhaseTransition}, we do this using the time-dependent variational principle (TDVP) \cite{haegeman2011time, haegeman2016unifying, vanderstraeten2019tangent} which involves solving a set of ordinary differential equations, for the tensors $\mathcal{M}$, which are obtained by projecting the wavefunction onto a manifold of bond dimension $\chi_{\max}$. We start with an initial state in the form of Eq.~\eqref{eq:MatrixProductState} and compute the time evolution using TEBD or TDVP. After each time step, we truncate the resultant MPS using singular value decomposition to bond dimension $\chi_{\max}$, which parameterizes the simulation of time evolution. In the following, we will drop the subscript to mean $\chi \equiv \chi_{\max}$.

The TEBD and TDVP methods are described in detail in the literature~\cite{vidal2003efficient, vidal2004efficient, daley2004time, schollwock2011density, paeckel2019timeevolution, haegeman2011time, haegeman2016unifying}. We describe details of our implementation below.
%
%%%%%%%%%%%%%%%%%%%%%%%%%%%%%%%%%%%%%%%%%%%%%%%%%%%%%%%%%%%%%%%%%%%%%%%%%%%%%%%%
\subsubsection{Time evolving block decimation}
\label{app:TEBD}
%%%%%%%%%%%%%%%%%%%%%%%%%%%%%%%%%%%%%%%%%%%%%%%%%%%%%%%%%%%%%%%%%%%%%%%%%%%%%%%%
%
Time-evolving block decimation (TEBD) is a powerful method to approximate the time evolution operator generated by local Hamiltonians, with a few neighboring interactions, on a 1D lattice \cite{vidal2003efficient, vidal2004efficient, daley2004time, schollwock2011density, paeckel2019timeevolution}. The approximation involves, approximating the time evolution operator, $\exp\left(-\mathrm{i} t H \right)$.
%, which is an order-$2n$ tensor, for $n$ degrees of freedom. 
The approximation involves a Trotter-Suzuki decomposition of the time evolution operator into its local terms.

The simulations for the nearest neighbor TFIM and SFIM in Sec.~\ref{sec:CorrelationLengthFromQuench} and Sec.~\ref{sec:RoleChaos} are performed using the TEBD algorithm~\cite{vidal2003efficient, vidal2004efficient, daley2004time}. The Hamiltonian of the SFIM considered in Sec.~\ref{sec:RoleChaos} is given by
\begin{equation} 
\begin{aligned}
    H = &
    - J_0 \sum_{\ell} \sigma^z_{\ell} \sigma^z_{\ell + 1}
    - B \sum_{\ell} \left( \cos(\theta)
    \sigma^z_{\ell} + \sin(\theta) \sigma^x_{\ell} \right).
    \label{eq:HamiltonianSFIMAppendix}
\end{aligned}
\end{equation}
This Hamiltonian can be written as a sum over even $l$ and a sum over odd $l$,
\begin{equation}
    H = H_{\mathrm{even}} + H_{\mathrm{odd}}
    ,
\end{equation}
where $H_{\mathrm{even}}$ corresponds to terms with an even value of $l$ and $H_{\mathrm{odd}}$ corresponds to terms with an odd value of $l$ as
\begin{equation}
\begin{aligned}
    H_{\mathrm{even}}
    &
    = \sum_{l \in \mathrm{even}} h_{l}
    \\
    H_{\mathrm{odd}}
    &
    = \sum_{l \in \mathrm{odd}} h_{l}
    ,
\end{aligned}
\end{equation}
where $h_{l}$ is the two-local Hamiltonian acting on sites $l$ and $l+1$
\begin{equation}
\begin{aligned}
    h_{l} =
    &
    - J \sigma^z_{l} \sigma^z_{l+1}
    \\ &
    - \frac{1}{2} B\cos(\theta) \left(\sigma^z_l + \sigma^z_{l+1} \right)
    \\ &
    - \frac{1}{2} B\sin(\theta) \left(\sigma^x_l + \sigma^x_{l+1} \right)
    ,
\end{aligned}
\end{equation}
where the factor of $1/2$ is introduced as the one-local terms, $\sigma^{z}_{l+1}$ and $\sigma^{x}_{l+1}$ are present in both $h_{l}$ and $h_{l+1}$. The only exception to this are the boundary terms in $h_0$ and $h_n$. The overlap between $h_{l}$ and $h_{l+1}$ make them not commute, $[h_{l}, h_{l+1}] \neq 0$. Therefore, $H_{\mathrm{even}}$ does not commutes with $H_{\mathrm{odd}}$. However, $H_{\mathrm{even}}$ and $H_{\mathrm{odd}}$ are each a sum of commuting terms.

With these noncommuting terms, we use a fourth order Trotter-Suzuki product formula with a time step $\Delta t = 0.01$ and vary the system size $n$ and bond dimension $\chi$.

%
%%%%%%%%%%%%%%%%%%%%%%%%%%%%%%%%%%%%%%%%%%%%%%%%%%%%%%%%%%%%%%%%%%%%%%%%%%%%%%%%
\subsubsection{Time-dependent variational principle}
\label{app:TDVP}
%%%%%%%%%%%%%%%%%%%%%%%%%%%%%%%%%%%%%%%%%%%%%%%%%%%%%%%%%%%%%%%%%%%%%%%%%%%%%%%%
%
The time-dependent variational principle (TDVP) is another way of approximating the dynamics of MPS by reducing the many-body time evolution to a series of local time evolution problems \cite{haegeman2011time, haegeman2016unifying, paeckel2019timeevolution}. TDVP involves constraining the time evolution to a specific manifold of MPS with a given bond dimension. TDVP applied to MPS is a convenient method for calculating the dynamics of long-range interacting models in 1D \cite{haegeman2016unifying, paeckel2019timeevolution}. It is also convenient for approximate time evolution which respects the conservation of quantities that are symmetries of the Hamiltonian.

In Sec.~\ref{sec:DynamicalQuantumPhaseTransition}, we use the two-site variant, 2TDVP by considering the action of the Hamiltonian on a pair of neighboring tensors in the MPS \cite{haegeman2016unifying}. This gives a differential equation for every pair of neighbors. While the solution of these differential equations does not project each tensor onto a fixed bond dimension manifold, subsequent SVD and truncation do \cite{haegeman2016unifying}. We use the implementation in the software library \texttt{TimeEvoMPS.jl}~\cite{TimeEvoMPS} using \texttt{ITensor.jl} version 0.3~\cite{ITensor, ITensor-r0.3}.

We express the Hamiltonian in Eq.~\ref{eq:HamiltonianPowerLawTFIM} as a matrix product operator (MPO). Different simulations are performed with a time step $\Delta t = 0.01$ and varying the system size $n$ and bond dimension $\chi$.

%
%%%%%%%%%%%%%%%%%%%%%%%%%%%%%%%%%%%%%%%%%%%%%%%%%%%%%%%%%%%%%%%%%%%%%%%%%%%%%%%%
\section{Additional data for extracting critical point using minimum}
\label{app:DQPTCriticalPointMinimum}
\begin{figure}[htbp] % 
   \centering
   \subfigure[\label{fig:DQPTParabola2}]{\includegraphics[width=0.48\textwidth]{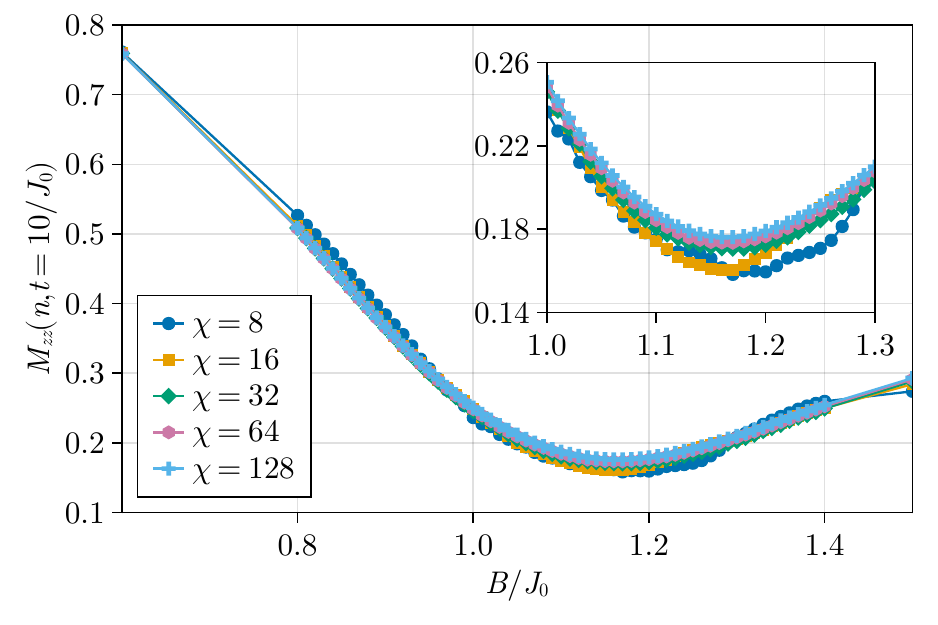} } 
   \subfigure[\label{fig:DQPTParabolaFit2}]{\includegraphics[width=0.48\textwidth]{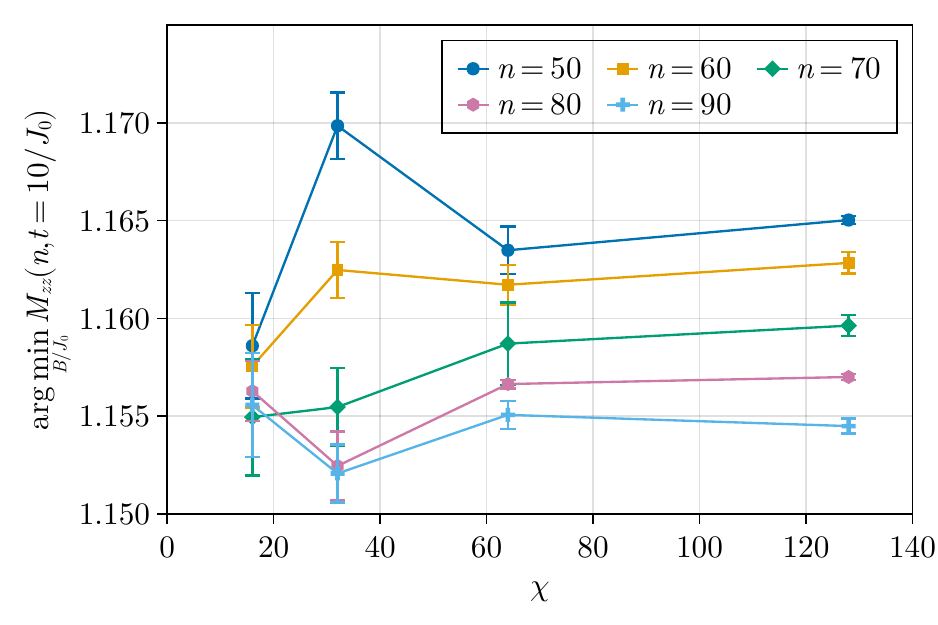} }
   \caption{ (a) $M_{zz}$ versus $B/J_0$ for $n=50$, $\alpha = 1.5$, $J_0 t = 10$ and different bond dimension $\chi$. (b) Fit of the location of the minimum of $M_{zz}$ with respect to $B/J_0$ for $\alpha = 1.5$, $J_0 t = 10$, shown for different system sizes $n$ and bond dimensions $\chi$. These plots show the robustness to MPS truncation in the estimation of the minimum in the order parameter, even for a relatively large number of spins and longer times.}
   \label{fig:DQPTmin2}
\end{figure}

%%%%%%%%%%%%%%%%%%%%%%%%%%%%%%%%%%%%%%%%%%%%%%%%%%%%%%%%%%%%%%%%%%%%%%%%%%%%%%%%
%
In Sec.~\ref{sec:DQPTmin} of the main text we used the the minimum in $M_{zz}$ at $J_0 t = 5$ to identify the location of the DQPT. Here we repeat this analysis using a later time $J_0 t = 10$. In Fig.~\ref{fig:DQPTParabola2} and Fig.~\ref{fig:DQPTParabolaFit2} we show the behavior of $M_{zz}$ as a function of $B/J_0$ at this later time. The variation for the location of the minimum between different system size is smaller than observed for $J_0 t = 5$ in the main text, however we observe similar behavior: there is a very small dependence on the bond dimension, and the location of the minimum steadily decreases with system size for all bond dimensions $\chi \geq 32$.
%
%%%%%%%%%%%%%%%%%%%%%%%%%%%%%%%%%%%%%%%%%%%%%%%%%%%%%%%%%%%%%%%%%%%%%%%%%%%%%
\section{More finite-size scaling results}
\label{app:FiniteSizeScalingMz}
%%%%%%%%%%%%%%%%%%%%%%%%%%%%%%%%%%%%%%%%%%%%%%%%%%%%%%%%%%%%%%%%%%%%%%%%%%%%%
%

\begin{figure*}[htbp] % 
   \centering
   \subfigure[]
   {\includegraphics[width=0.42\textwidth]{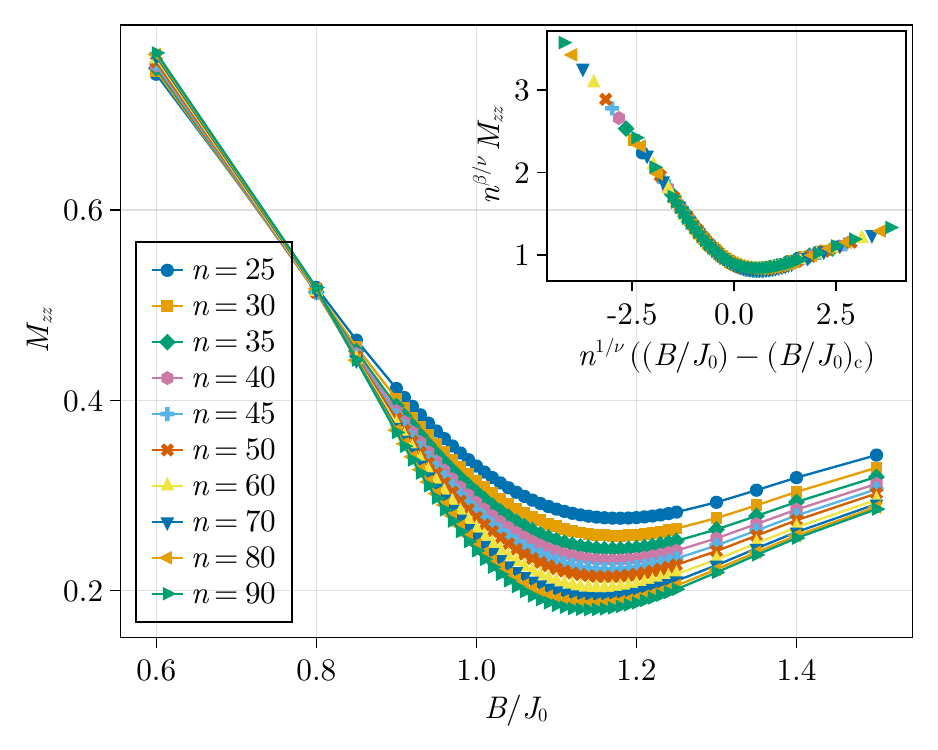} \label{fig:DataCollapseSpaceExample}}
   \subfigure[]
   {\includegraphics[width=0.42\textwidth]{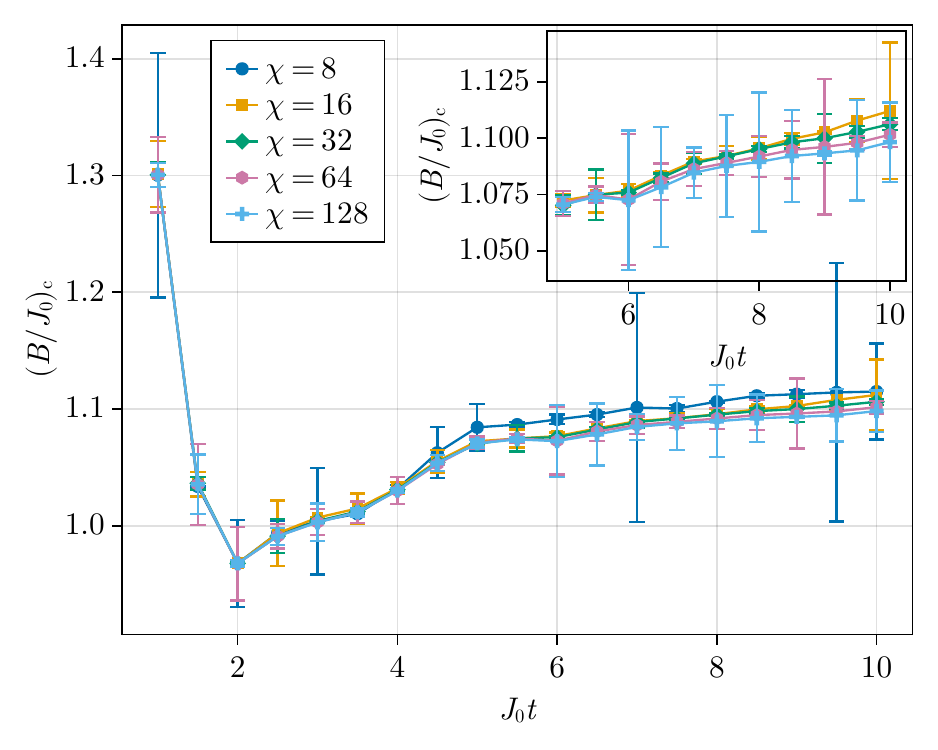} \label{fig:DataCollapseSpaceBCritical}}
   \subfigure[]
   {\includegraphics[width=0.42\textwidth]{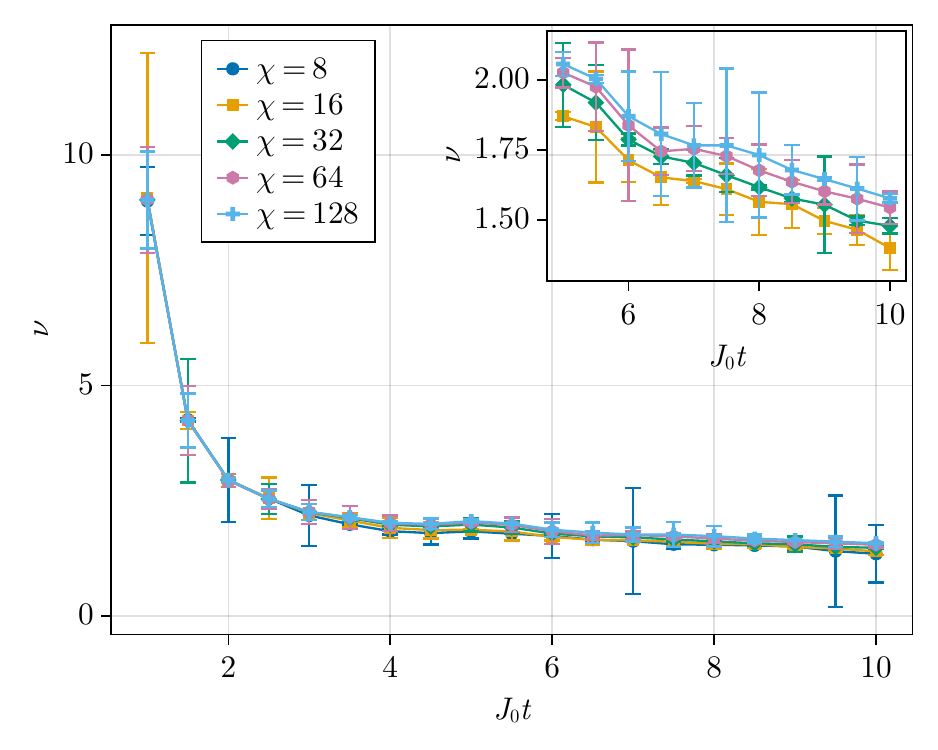} \label{fig:DataCollapseSpaceNu}}
   \subfigure[]
   {\includegraphics[width=0.42\textwidth]{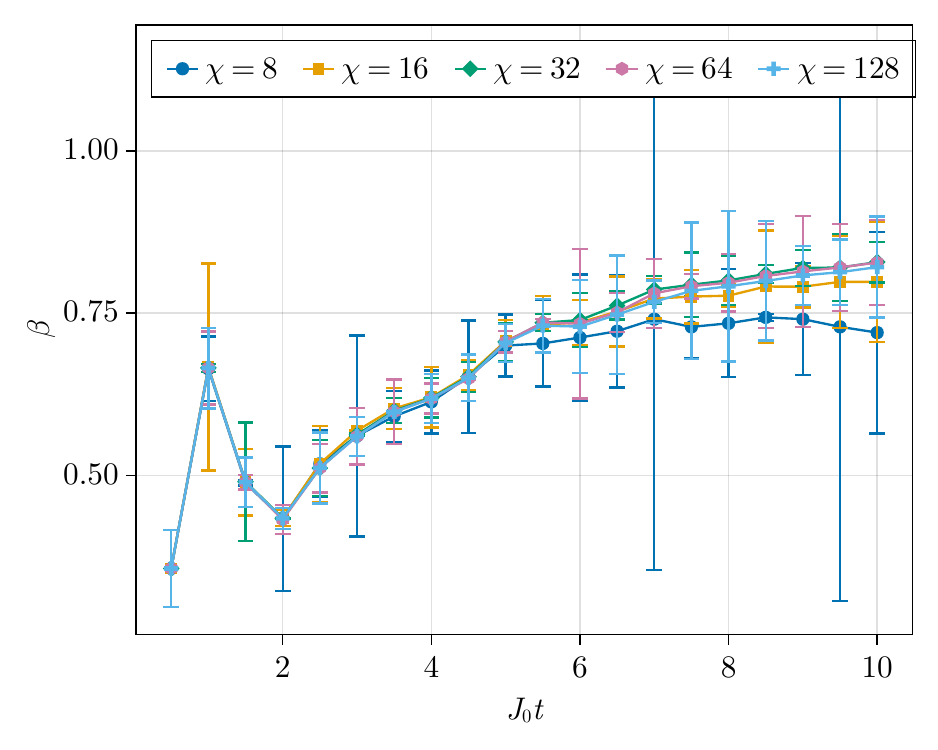} \label{fig:DataCollapseSpaceZeta}}
   \caption{ (a) Finite-size scaling analysis for order parameter $M_{zz}$ for simulations using $\alpha =1.5, J_0 t = 5, \chi = 128, J_0 \Delta t = 10^{-2}$. Using the data collapse method, the extracted critical parameters for these simulations parameters from this simulation are $\left(B/J_0\right)_{\mathrm{c}} = 1.071 \pm 0.003, \nu = 2.05 \pm 0.04, \beta = 0.70 \pm 0.03$. Using the data collapse method, the extracted critical point (b) and critical exponents (c), (d) as a function of dimensionless time, $J_0t$, for different bond dimensions $\chi \in \{8, 16, 32, 64, 128\}$. Insets show estimates for $J_0t \gtrsim 5.0$, which are more consistent. The estimates of the critical point and critical exponents are approximately the same across different bond-dimensions.}
	\label{fig:DataCollapseParametersSpace}
\end{figure*}

To supplement the results of finite-size scaling (FSS) in the main text, here we present results on applying FSS to $M_{zz}$ to estimate the critical point and critical exponents of the DQPT we considered in Sec.~\ref{sec:DynamicalQuantumPhaseTransition}. As in the main text we use the data collapse method \cite{New1999,Bin2010}, assuming an equilibrium FSS ansatz
\begin{equation}
  M_{zz}(n) = n^{-\beta/\nu} F\left(n^{1/\nu}
    \left(\frac{B}{J_0} -
    \left(\frac{B}{J_0}\right)_{\mathrm{c}}\right); J_0 t \right),
    \label{eq:FiniteSizeScalingSpaceMzzAnsatz}
\end{equation}
with a universal scaling function $F$ (not necessarily the same function $f$ used for $M_{z}$ in the main text). We show an example of the results of the data collapse method in Fig.~\ref{fig:DataCollapseSpaceExample}. As with $M_{z}$ in Sec.~\ref{sec:DynamicalQuantumPhaseTransition}, we obtain estimates of the critical parameters. In particular we show critical point estimates in Fig.~\ref{fig:DataCollapseSpaceBCritical}, and the estimates of critical exponent $\nu$ and $\beta$ in Fig.~\ref{fig:DataCollapseSpaceNu} and Fig.~\ref{fig:DataCollapseSpaceZeta}, respectively, for different times $t$ and bond-dimension $\chi$. The function $F$ behaves as $F(x) \sim |x|^{\beta}$ as $|x| \to \infty$.

Similar to $M_{z}$ in the main text, for short duration quenches $J_0t \lesssim 2$ the FSS results for $M_{zz}$ are inconclusive, suggesting that local equilibration has not occurred. For longer quenches $J_0t \gtrsim 5$, the estimates are stable, but with a temporal drift. The rapid equilibration which we observed in Sec.~\ref{sec:DynamicalQuantumPhaseTransition} enables us to observe characteristics of the phase transition at short times $J_0 t \ll n$. 

We note that this data collapse and the one for $M_{z}$ in Sec.~\ref{sec:DynamicalQuantumPhaseTransition} find a finite $\nu$, which suggests a diverging correlation length  at the critical point in the thermodynamic limit. However, Lieb-Robinson bounds for the $\alpha = 1.5$ power-law interacting system limit the rate of growth of correlations \cite{Tran2019, Chen2019}. 
We expect the drift in the estimates of the critical exponents $\nu$ and $\beta$ with time to be a reflection of this growth. 
An understanding of the nature of the evolving divergence and it relationship with Lieb-Robinson bounds and the thermal phase transition is an avenue for future work.

Even with the temporal drift, the estimates of the critical point $(B/J_0)_{\mathrm{c}}$ and critical exponents $\nu, \beta$ are approximately the same across different bond-dimensions. Specifically estimates at $\chi = 16$ are quite close to the estimates at $\chi = 128$. Therefore, many-body entanglement beyond a R{\'e}nyi-0 entropy of $\mathcal{S}_0 = \ln(16)$ plays a negligible role in the order parameter $M_{zz}$ in this phase transition, just as they did with the order parameter $M_{z}$.

\begin{figure*}[htbp] % 
   \centering
   \subfigure[]
   {\includegraphics[width=0.42\textwidth]{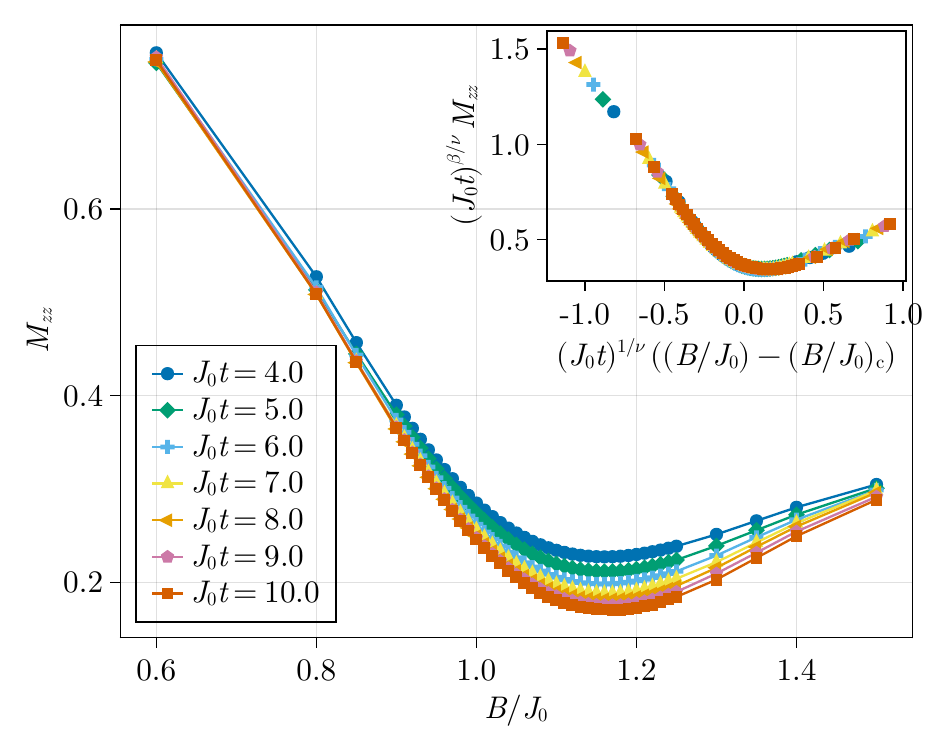} \label{fig:DataCollapseTimeExample}}
   \subfigure[]
   {\includegraphics[width=0.42\textwidth]{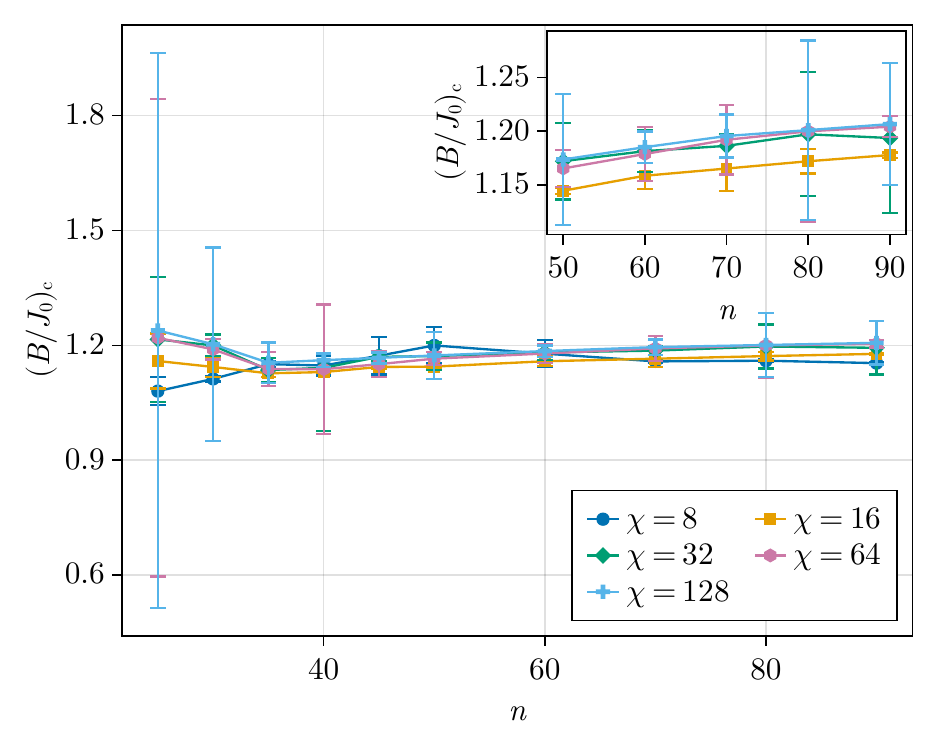} \label{fig:DataCollapseTimeBCritical}}
   \subfigure[]
   {\includegraphics[width=0.42\textwidth]{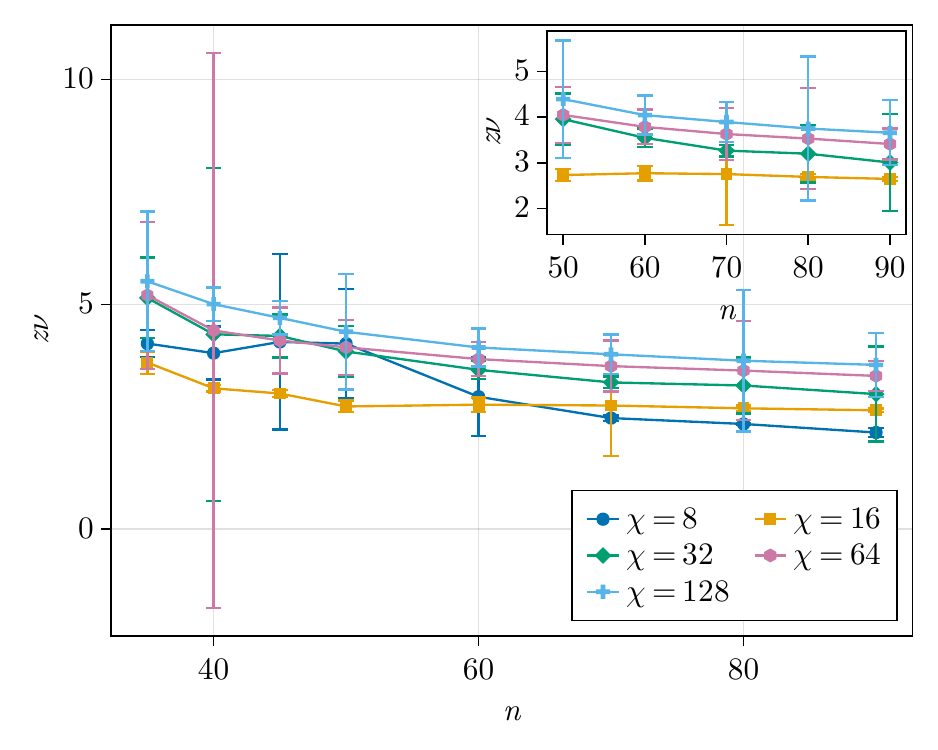} \label{fig:DataCollapseTimeNu}}
   \subfigure[]
   {\includegraphics[width=0.42\textwidth]{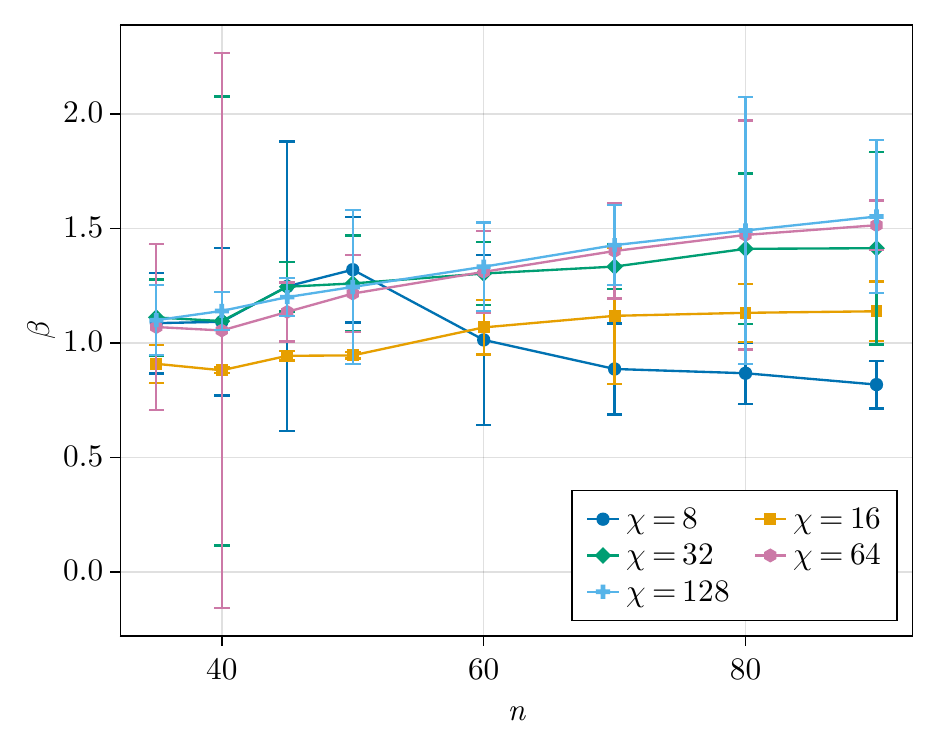} \label{fig:DataCollapseTimeZeta}}
   \caption{ (a) Finite-size scaling analysis for order parameter $M_{zz}$ for simulations using $\alpha =1.5, n = 50, \chi = 32, \Delta t = 10^{-2}$. Using the data collapse method, the extracted critical parameters for these simulations parameters from this simulation are $\left(B/J_0\right)_{\mathrm{c}} = 1.0979 \pm 0.0007, z\nu = 2.8 \pm 0.3, \beta = 0.85 \pm 0.09$. Using the data collapse method, the extracted critical point (b) and critical exponents (c), (d) as a function of system size $n$, for different bond dimensions $\chi \in \{8, 16, 32, 64, 128\}$. Insets show estimates for $n \geq 50$, which are more consistent. The estimates of the critical point and critical exponents are approximately the same across different bond-dimensions.} \label{fig:DataCollapseParametersTime}
\end{figure*}

We now consider a non-equilbrium FSS ansatz of the form
\begin{equation}
    M_{zz}(t) = (J_0 t)^{-\beta/z\nu} G\left((J_0 t)^{1/z\nu}
    \left(\frac{B}{J_0} -
    \left(\frac{B}{J_0}\right)_{\mathrm{c}}\right); n\right),
    \label{eq:FiniteSizeScalingTimeMzzAnsatz}
\end{equation}
for FSS analysis in the temporal direction. $G$ is a scaling function (not necessarily the same function $g$ used for $M_{z}$) and $z$ is a dynamical critical exponent. We give an example of the results of the data collapse method in Fig.~\ref{fig:DataCollapseTimeExample}. Performing this analysis for simulations for different system sizes $n$ and bond-dimension $\chi$, we get different estimates of the critical parameters. In particular we show the critical point estimates in Fig.~\ref{fig:DataCollapseTimeBCritical}, critical exponent $z \nu$ in Fig.~\ref{fig:DataCollapseTimeNu}, and critical exponent $\beta$ in Fig.~\ref{fig:DataCollapseTimeZeta}. The function $G$ behaves as $G(x) \sim |x|^{\beta}$ as $|x| \to \infty$. As in our previous cases, the estimates of the critical properties are approximately the same across different bond-dimensions. 

\bibliography{refs}

\end{document}